\documentclass[preprint2,times,tighten]{aastex6}
\pdfoutput=1 
\usepackage{amsmath,amstext}
\usepackage[figure,figure*]{hypcap}
\usepackage{hyperref}
\usepackage{newtxmath} 
\usepackage{threeparttable}

\slugcomment{\it Paper II --- 2017 August}

\def\deg{\ifmmode {^{\circ}}\else {$^\circ$}\fi}
\def\kms{\ifmmode {\rm\,km\,s^{-1}}\else
    ${\rm\,km\,s^{-1}}$\fi}
\def\ergcm2s{\ifmmode {\rm\,erg\,cm^{-2}\,s^{-1}}\else
    ${\rm\,erg\,cm^{-2}\,s^{-1}}$\fi}
\def\ergAcm2s{\ifmmode {\rm\,erg\,cm^{-2}\,s^{-1}\,\AA^{-1}}\else
    ${\rm\,erg\,cm^{-2}\,s^{-1}\,\AA^{-1}}$\fi}
\def\ergs{\ifmmode {\rm\,erg\,s^{-1}}\else
    ${\rm\,erg\,s^{-1}}$\fi}
\def\kmsMpc{\ifmmode {\rm\,km\,s^{-1}\,Mpc^{-1}}\else
    ${\rm\,km\,s^{-1}\,Mpc^{-1}}$\fi}

\def\oii{[\ion{O}{2}] $\lambda$3727}

\def\spose#1{\hbox to 0pt{#1\hss}}
\def\simlt{\mathrel{\spose{\lower 3pt\hbox{$\mathchar"218$}}
     \raise 2.0pt\hbox{$\mathchar"13C$}}}
\def\simgt{\mathrel{\spose{\lower 3pt\hbox{$\mathchar"218$}}
     \raise 2.0pt\hbox{$\mathchar"13E$}}}
\def\pz{P($z|C$)}
\def\rhoc{$\rho$$(C)$}
\def\phz{photo-$z$}
\def\sz{spec-$z$}
\def\nz{$N(z)$}
\def\e{\emph{Euclid}}
\def\w{\emph{WFIRST}}

\def\deg{$^{\rm o}$}

\def\arcsec{\ifmmode '' \else $''$\fi}

\def\arcsecpoint{\ifmmode ''\!. \else $''\!.$\fi}

\def\kms{\ifmmode {\rm km\ s}^{-1} \else km s$^{-1}$\fi}
\def\Msun{\ifmmode {\rm M}_{\odot} \else M$_{\odot}$\fi}
\def\Lsun{\ifmmode {\rm L}_{\odot} \else L$_{\odot}$\fi}
\def\Zsun{\ifmmode {\rm Z}_{\odot} \else Z$_{\odot}$\fi}

\def\ergscm2{ergs\,s$^{-1}$\,cm$^{-2}$}
\def\icm3{{\rm cm}^{-3}}
\def\icm2{{\rm cm}^{-2}}
\def\qo{\ifmmode q_{\rm o} \else $q_{\rm o}$\fi}
\def\Ho{\ifmmode H_{\rm o} \else $H_{\rm o}$\fi}
\def\ho{\ifmmode h_{\rm o} \else $h_{\rm o}$\fi}

\def\vFWHM{\ifmmode v_{\mbox{\tiny FWHM}} \else
            $v_{\mbox{\tiny FWHM}}$\fi}

\def\gtorder{\mathrel{\raise.3ex\hbox{$>$}\mkern-14mu
             \lower0.6ex\hbox{$\sim$}}}
\def\ltorder{\mathrel{\raise.3ex\hbox{$<$}\mkern-14mu
             \lower0.6ex\hbox{$\sim$}}}
\def\proptwid{\mathrel{\raise.3ex\hbox{$\propto$}\mkern-14mu
             \lower0.6ex\hbox{$\sim$}}}

\begin{document}

\title{The Complete Calibration of the Color-Redshift Relation (C3R2) Survey:\\
Analysis and Data Release 2}

\author{Daniel C. Masters\altaffilmark{1}, Daniel K. Stern\altaffilmark{1}, Judith G. Cohen\altaffilmark{2}, Peter L. Capak\altaffilmark{3}, S. Adam Stanford\altaffilmark{4}, Nina Hernitschek\altaffilmark{2}, Audrey Galametz\altaffilmark{5}, Iary Davidzon\altaffilmark{6}, Jason D. Rhodes\altaffilmark{1,11}, Dave Sanders\altaffilmark{7}, Bahram Mobasher\altaffilmark{8}, Francisco Castander\altaffilmark{9,10}, Kerianne Pruett\altaffilmark{4}, Sotiria Fotopoulou\altaffilmark{12}
}

\altaffiltext{1}{
Jet Propulsion Laboratory, California Institute of Technology, Pasadena,
CA 91109, USA}
\altaffiltext{2}{
California Institute of Technology, Pasadena,
CA 91125, USA}
\altaffiltext{3}{
Spizter Science Center, Pasadena,
CA 91125, USA}
\altaffiltext{4}{Department of Physics,
University of California, Davis,
CA 95616, USA}
\altaffiltext{5}{Department of Astronomy, University of Geneva, 1205, Versoix, Switzerland}
\altaffiltext{6}{
Infrared Processing and Analysis Center, California Institute of Technology, Pasadena,
CA 91125, USA}
\altaffiltext{7}{
Institute for Astronomy, University of Hawaii, 96822, USA }
\altaffiltext{8}{Department of Physics and Astronomy, University of California, Riverside, CA 92521, USA}
\altaffiltext{9}{Institut d’Estudis Espacials de Catalunya (IEEC), 08034 Barcelona, Spain}

\altaffiltext{10}{Institute of Space Sciences (ICE, CSIC), Campus UAB, Carrer de Can Magrans, s/n, 08193 Barcelona, Spain}
\altaffiltext{11}{Kavli Institute for the Physics and Mathematics of the Universe (IPMU), Tokyo, Japan}
\altaffiltext{12}{Center for Extragalactic Astronomy, Department of Physics, Durham University, Durham, UK}

\begin{abstract} 

The Complete Calibration of the Color-Redshift Relation (C3R2) survey is a multi-institution, multi-instrument survey that aims to map the empirical relation of galaxy color to redshift to $i$$\sim$24.5~(AB), thereby providing a firm foundation for weak lensing cosmology with the Stage IV dark energy missions \e\ and \w. Here we present 3171 new spectroscopic redshifts obtained  in the 2016B and 2017A semesters with a combination of DEIMOS, LRIS, and MOSFIRE on the Keck telescopes\footnote{Redshifts and spectra released by C3R2 to date can be found at \url{https://sites.google.com/view/c3r2-survey/home} and \url{https://koa.ipac.caltech.edu/Datasets/C3R2}.}. The observations come from all of the Keck partners: Caltech, NASA, the University of Hawaii, and the University of California. Combined with the 1283 redshifts published in DR1, the C3R2 survey has now obtained and published 4454 high quality galaxy redshifts. We discuss updates to the survey design and provide a catalog of photometric and spectroscopic data. Initial tests of the calibration method performance are given, indicating that the sample, once completed and combined with extensive data collected by other spectroscopic surveys, should allow us to meet the cosmology requirements for \e, and make significant headway toward solving the problem for \w. We use the full spectroscopic sample to demonstrate that galaxy brightness is weakly correlated with redshift once a galaxy is localized in the \e\ or \w\ color space, with potentially important implications for the spectroscopy needed to calibrate redshifts for faint \w\ and LSST sources. 

\end{abstract}


\section{Introduction}

Stage IV cosmology missions (LSST, \e, and \w) will use deep imaging of billions of galaxies in filters spanning the optical to the near-infrared to estimate  photometric redshifts (\phz s) for weak lensing cosmology. While high-quality photo-$z$s are also crucial for many other investigations, the \phz\ requirements for cosmology are particularly stringent. Weak lensing cosmology  requires \textit{unbiased} redshift estimates \citep{Ma06,Huterer06}, at a level which is not possible with classical \phz\ estimation techniques such as spectral template fitting. For \e\ and \w, the requirement is usually given that the mean redshift $\langle z \rangle$ of galaxies in $\sim$10 shear bins must be known to better than 0.2\% (that is, $\Delta \langle z \rangle \leq 0.002 (1+\langle z \rangle)$ for each bin). Numerous tests have shown that this level of accuracy is not achieved with traditional algorithms and realistic training samples. The bias requirement, in particular, makes spectroscopic calibration samples  necessary  for the success of these missions \citep{Newman13}.

Stage III cosmology surveys currently underway also face the \phz\ estimation challenge, and constitute an important testbed for the \phz\ techniques to be employed in the Stage IV experiments. These include the Kilo-Degree Survey (KiDS; \citealp{deJong15, deJong17, Hildebrandt17}), the Dark Energy Survey (DES; \citealp{des17,Troxel18}), and the Hyper-Suprime Cam (HSC) Survey \citep{HSC18}. A variety of techniques have been employed in these surveys to constrain redshift distributions, \nz, of galaxies in shear bins -- from template fitting (e.g., \citealp{Benitez00,Brammer08,Ilbert09}), to machine learning techniques based on training samples (e.g., \citealp{Collister04,Carrasco13a}), and clustering redshifts which use the spatial distribution of overlapping \sz\ samples to infer the \nz\ distribution of the photometric sample (e.g., \citealp{Newman08,Menard13,McQuinn13, Schmidt13, Morrison17}). Re-weighting of the spectroscopic sample to better match the photometric sample, as described in \citet{Lima08}, has also been used. It is generally agreed that a key limitation  is the need for a fully representative training sample of \sz s that explores the range of galaxy properties present in the surveys.

The Complete Calibration of the Color-Redshift Relation (C3R2) survey (\citealp{Masters17}, hereafter M17) was initiated in response to this need, with the goal of mapping the empirical relation between galaxy redshift and color to the \e\ depth of $i$$\sim$24.5~(AB). The survey strategy is based on the fact that the manifold of  observed galaxy colors to a given survey depth is both \textit{limited} and \textit{measurable}. Moreover, we make the (testable) assumption that there exists a well-defined, mostly non-degenerate relation between a galaxy's position in multi-color space and its redshift, which can be discovered empirically. Uncovering this \pz\ relation is the goal of the C3R2 survey.

The C3R2 survey strategy follows the method outlined in \citet{Masters15} (hereafter M15). M15 illustrated the use of an unsupervised manifold learning algorithm, the \emph{self-organizing map} (SOM; \citealp{Kohonen82}), to map the color distribution of galaxies in the high-dimensional color space (\emph{u-g, g-r, ..., J-H}) anticipated for \e\ and \w\ \phz\ estimation. This high-dimensional mapping allows us to directly determine which parts of galaxy color space are well sampled with existing spectroscopy and which are not, thus letting us focus spectroscopic calibration effort on those regions which are the least constrained. Essentially, we seek to obtain the \textit{minimum} additional spectroscopy needed to build a representative training sample, such that direct inference of the \pz\ relation can be made sufficiently accurate to meet the cosmology requirements.  

In M17 we presented the results of the first five nights of C3R2 observations taken in the 2016A semester. Here we present results obtained in the 2016B and 2017A semesters, comprising 23.5 Keck nights, of which $\sim$14 had good observing conditions. These nights were allocated as part of a multi-institution effort to solve the redshift calibration problem for \e\ and make significant progress toward the \w\ calibration, with time coming from all Keck partners: Caltech (10 nights, PI J. Cohen), NASA (5 nights, PI D. Stern), University of Hawaii (6 nights, PI D. Sanders), and the University of California (2.5 nights, PI B. Mobasher). A parallel effort with the VLT (PI F. Castander) is also underway, the results of which will be presented in DR3. We refer the reader to M15 for background on the calibration approach and to M17 for more details on the C3R2 survey. 

This paper is structured as follows. In \S2 we describe the development of homogeneous photometric catalogs in \emph{ugrizYJHK$_{s}$} for multiple deep fields. In \S3 we describe updates to the survey strategy. In \S4 we describe the observations and data reductions of the 2016B/2017A observations. In \S5 we  present our spectroscopic results from 23.5 nights in 2016B/2017A, and discuss performance of the method. In \S6 we investigate the status of the calibration effort, and issues still to be addressed.

\begin{deluxetable*}{cccccc}
\tabletypesize{\normalsize}
\tablecaption{Overview of the deep fields targeted by C3R2 in DR2.\label{tab:fields}}
\tablewidth{0pt}
\tablehead{
\colhead{Field} & \colhead{Right Ascension} & \colhead{Declination} & \colhead{Area} & \colhead{Optical data} &\colhead{Near-IR data} \\
\colhead{} & \colhead{(J2000)} & \colhead{(J2000)} & \colhead{(deg$^2$)} & \colhead{(\textit{ugriz})} & \colhead{(\textit{YJHK$_{s}$})}
}
\startdata
VVDS-2h & 02h 26m & $-04^\circ$ 30$^{\prime}$ & $1.0$ & CFHTLS & VISTA \\
COSMOS & 10h 00m & +02$^\circ$ 12$^\prime$ & $2.0$ & CFHTLS & VISTA \\ 
EGS & 14h 19m & +52$^\circ$ 41$^\prime$ & $1.0$ & CFHTLS & CFHTLS-WIRDS\tnote{\textdagger}\textdagger \\ 
\enddata
    \begin{tablenotes}
    \item[\textdagger] \textdagger \textit{Y} band obtained from CFHT-WIRCAM observations separate from the WIRDS survey.
    \end{tablenotes}
\end{deluxetable*}

\section{Constructing a Uniform Color System for Multiple Deep Fields}

Conducting the C3R2 survey across multiple fields at different right ascensions lets us take full advantage of observing nights and helps mitigate the effect of cosmic variance in an individual field. However, the sensitive multi-color selection technique we employ requires very homogeneous photometry between fields. The fields that can be used for C3R2 targeting are thus limited to those with uniform, well-calibrated photometry spanning the optical to near-infrared (near-IR), to depths comparable to or deeper than the planned \emph{Euclid} survey. 

\begin{figure}[htb]
\centering
 \includegraphics[width=0.47\textwidth]{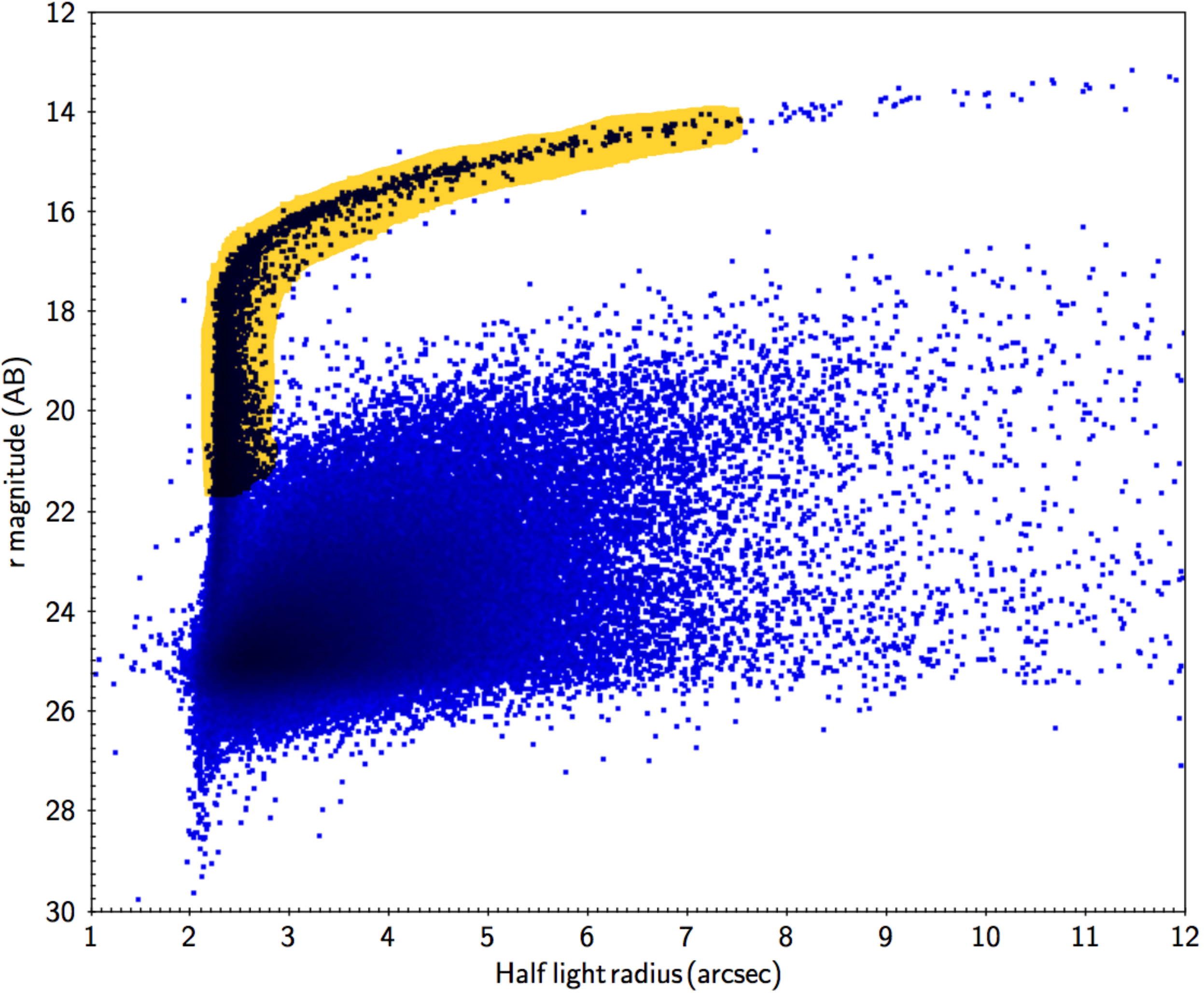} 
  \caption{The selection used to define the stellar locus for each field. The locus is clear in this plot of \emph{r} band magnitude vs. half-light radius; the highlighted region illustrates our selection. We select stars in the magnitude range $14\lesssim r \lesssim 22$ to test the uniformity of the color system.}
      \label{figure:cosmos_locus}
\end{figure}

We determined that the fields currently meeting this criterion are the Canada-France-Hawaii Telescope Legacy Survey\footnote{http://www.cfht.hawaii.edu/Science/CFHTLS/} (CFHTLS) deep fields (uniform MegaCam \emph{ugriz} imaging to AB point source depths of 26.3, 26.0, 25.6, 25.4, 25.0, respectively) which also have VISTA or CFHT-WIRCAM Deep Survey (WIRDS; \citealp{Bielby12}) near-IR imaging to the \e\ depth of \emph{YJH}$\sim$24 AB (5$\sigma$) (Table~1). These fields are VVDS-2h \citep{McCracken03, LeFevre04,Jarvis13}, COSMOS \citep{Scoville07,Capak07, McCracken12}, and EGS \citep{Davis07}.  In preparation for 2016B/2017A observations, we brought the near-IR photometry from these three deep fields onto a consistent $Euclid$/\emph{WFIRST}-like color system and collected existing redshifts from the literature\footnote{Note that the photometry from these fields is not corrected for galactic extinction. However, this correction is non-trivial because it is SED-dependent; moreover, the galactic extinction for these deep fields is low. Ultimately we will want to apply a galactic extinction correction based on the method developed in \citet{Galametz17}.}.

SXDS, CDFS, and GOODS-N were not included in our primary targeting because they lack the highly uniform photometry we believe is needed. However, all of these fields have extensive spectroscopy that can be added to the color-redshift relation with a rough color conversion to the \textit{Euclid}-like system. 

\subsection{Homogenizing the Near-Infrared Photometry}
Matching galaxy photometry from multiple fields at the desired level of precision is non-trivial because of the different instruments, filters, and photometry techniques employed in deep surveys. Fortunately the CFHTLS deep field imaging of VVDS-2h, COSMOS and EGS in CFHT MegaCam \emph{ugriz} is already highly uniform, which we confirmed by comparing the colors (measured with 2\arcsec\ aperture photometry) of stars in each field in various color-color diagrams.  

That left only the problem of deriving matched near-IR photometry for the fields. The CFHTLS-WIRDS survey \citep{Bielby12} obtained homogenized photometry for CFHT MegaCam optical and WIRCam near-IR (\textit{JHK$_{s}$}) imaging in these fields. However, subsequent VISTA imaging in VVDS-2h and COSMOS is both deeper than CFHTLS-WIRDS and includes \textit{Y} band. Our goal, therefore, was to bring the VVDS-2h and COSMOS VISTA observations onto the CFHTLS-WIRDS color  system. We now describe the steps taken to accomplish this.

\subsubsection{VVDS-2h}
To bring the VISTA data in VVDS-2h onto the CFHTLS-WIRDS system, we first needed to align the VISTA observations to the astrometric system of CFHTLS. We used the SCAMP and SWarp packages \citep{Bertin02, Bertin06} to compute and apply the needed astrometric correction to the VISTA imaging. The VISTA photometry was then extracted in an identical way as the CFHTLS-WIRDS data using the \citet{Bielby12} \emph{gri} detection images and SExtractor settings. A zero-point offset was then applied to the VISTA photometry to match WIRDS. No filter-dependent corrections were required because the filters of VISTA and CFHT-WIRCam are almost identical.

\subsubsection{COSMOS}
The COSMOS field has been the subject of extensive imaging spanning the electromagnetic spectrum, including from CFHTLS and UltraVISTA \citep{McCracken12}. The VISTA imaging from UltraVISTA is already on the same astrometric system as the CFHTLS data. As with VVDS-2h, we re-extracted the COSMOS UltraVISTA photometry using the same \emph{gri} detection images and SExtractor settings as the CFHTLS-WIRDS survey. A zero-point correction was derived and applied to match the UltraVISTA photometry to CHFTLS-WIRDS.

\subsubsection{EGS}
EGS has not been observed with VISTA, as it is too far north. Therefore, for this field we used the \textit{JHK$_{s}$} near-IR catalogs from CFHTLS-WIRDS directly. CHFTLS-WIRDS does not include \emph{Y} observations, so we downloaded archival \emph{Y} imaging from the CFHT Science Archive. We generated mosaics using SCAMP and SWarp on the output of the pipeline process data from the Canadian Astronomy Data Centre (CADC). We then extracted the photometry with SExtractor using the CFHTLS \textit{gri} detection image. Finally, we used the measured stellar locus in colors involving \textit{Y} band to force the reduced EGS \textit{Y} photometry to match the COSMOS data. Of the three fields, the color system for EGS is the least robust, due to the lower depth of the CFHTLS-WIRDS data compared with VISTA as well as the suboptimal \emph{Y} band imaging. 

\begin{figure*}[htb]
\centering
 \includegraphics[width=0.9\textwidth]{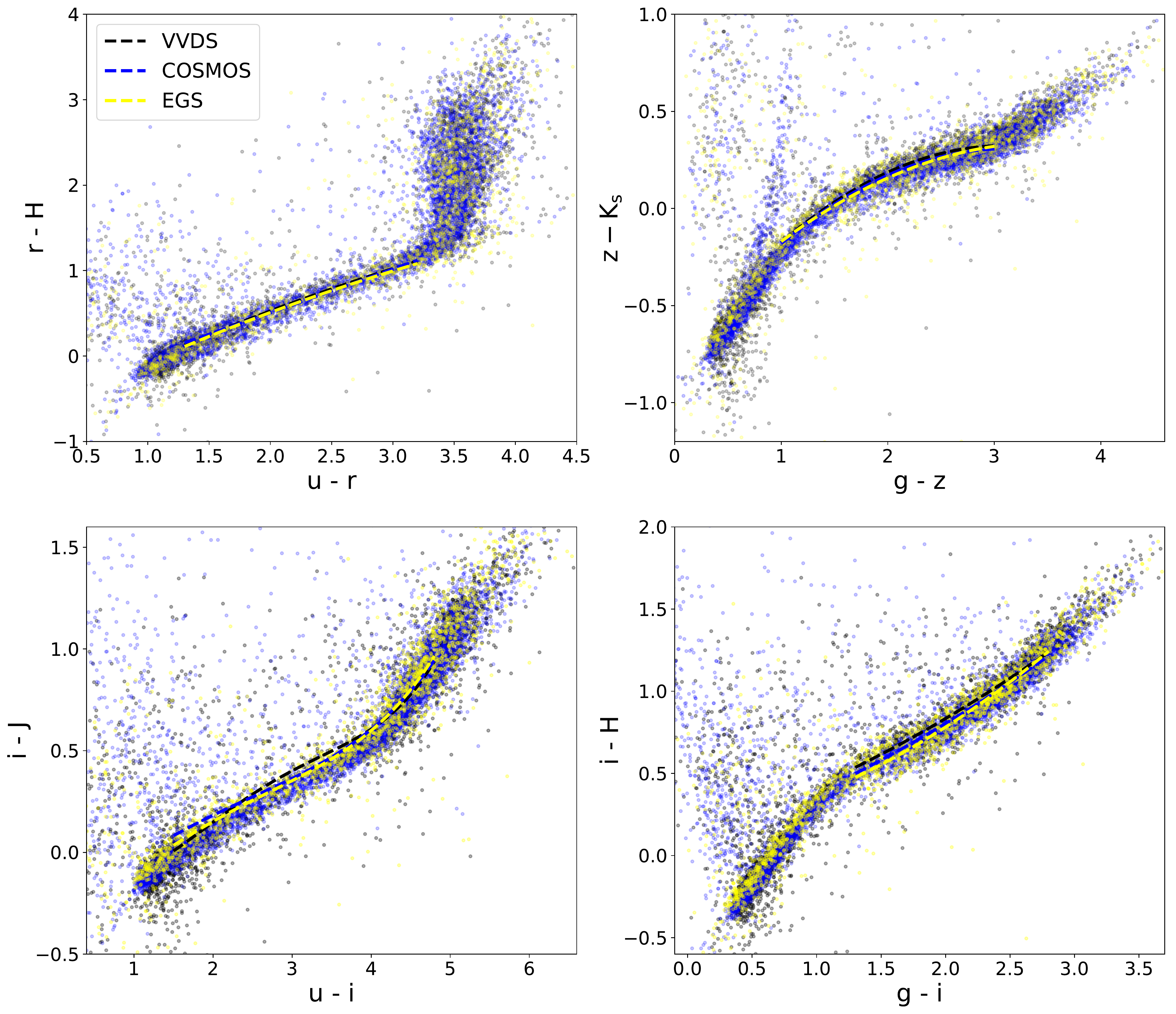} 
    
  \caption{Using stars selected from the three key fields (COSMOS, VVDS-2h, and EGS), we examine the stellar locus in different color-color plots spanning the optical to the near-IR, illustrating the overall excellent color agreement of the fields. This agreement is a necessary condition to employ a sensitive multi-color selection across different fields. In general the loci agree to $\lesssim$0.02 mag, as confirmed with the low-order polynomial fits shown. The largest residual offsets we find are between the VVDS-2h and COSMOS/EGS fields in the lower left plot (\emph{u-i} vs. \emph{i-J}), at the $\lesssim$0.05 magnitude level. The sources that scatter off the locus in these diagrams are primarily a combination of AGN and non-main sequence stars.}
      \label{figure:stellar_locus}
\end{figure*}

\begin{figure*}[htb]
\centering
 \includegraphics[width=0.9\textwidth]{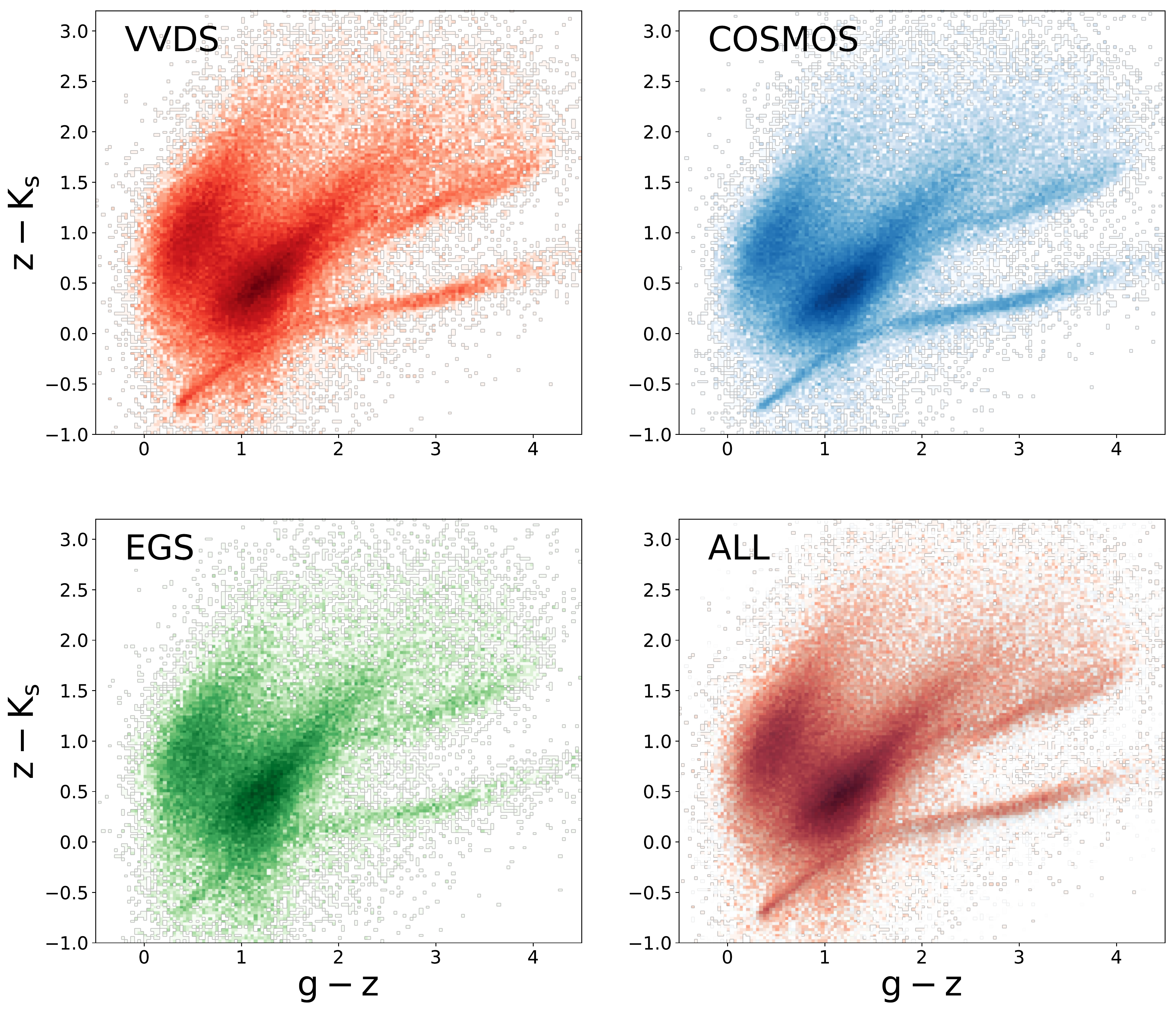} 
  \caption{Comparison of galaxy colors in the three C3R2-targeted fields in the \textit{gzK$_{s}$} diagram. This figure illustrates the overall close uniformity of galaxy colors (in addition to the stellar colors) between the three fields. In the bottom right figure we have overlaid the distributions with increasing transparency; essentially no significant difference is apparent.}
      \label{figure:gzK}
\end{figure*}

\begin{figure*}[htb]
\centering
  \begin{tabular}{@{}ccc@{}}
    \includegraphics[width=.32\textwidth]{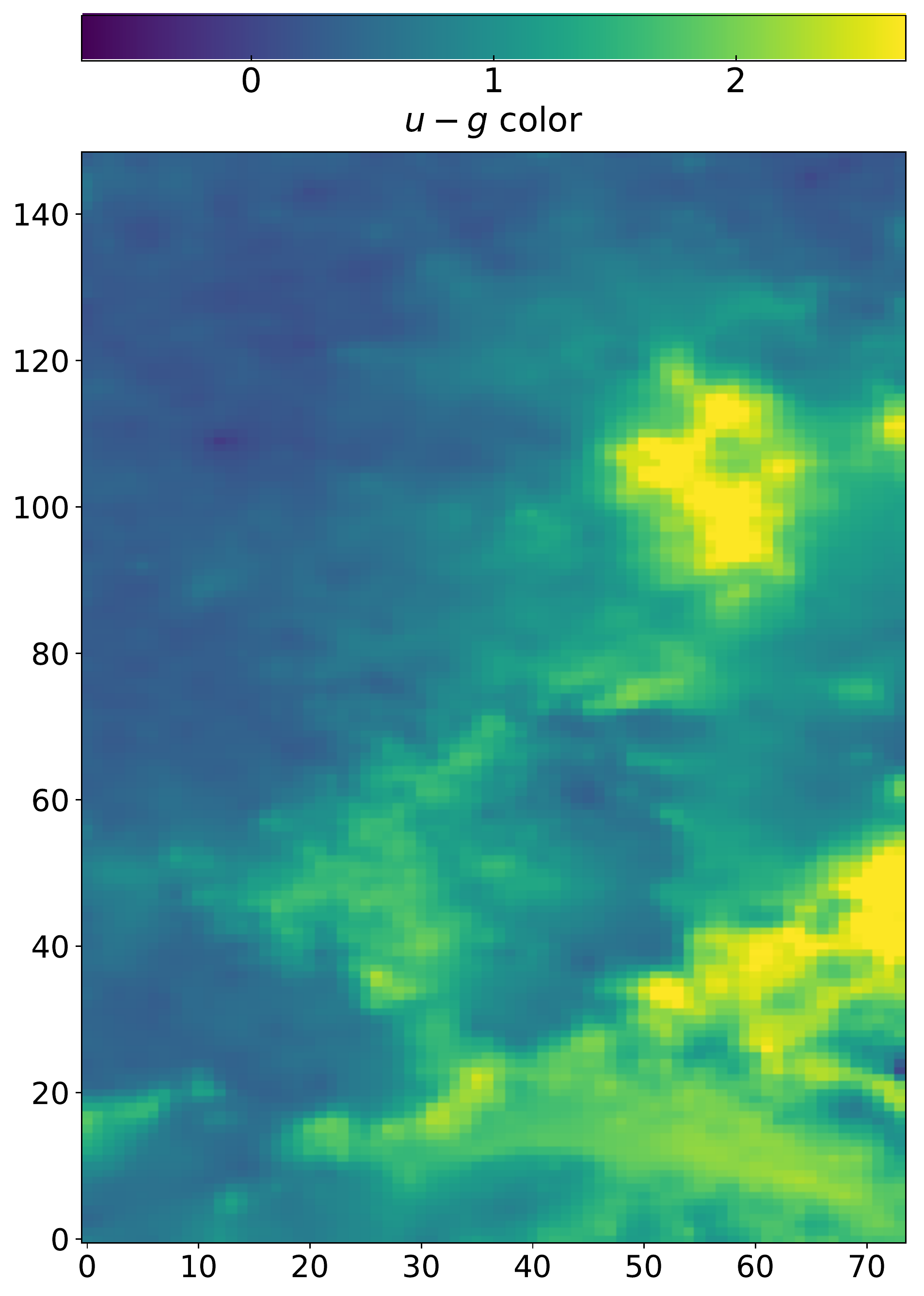} &
	 \includegraphics[width=0.32\textwidth]{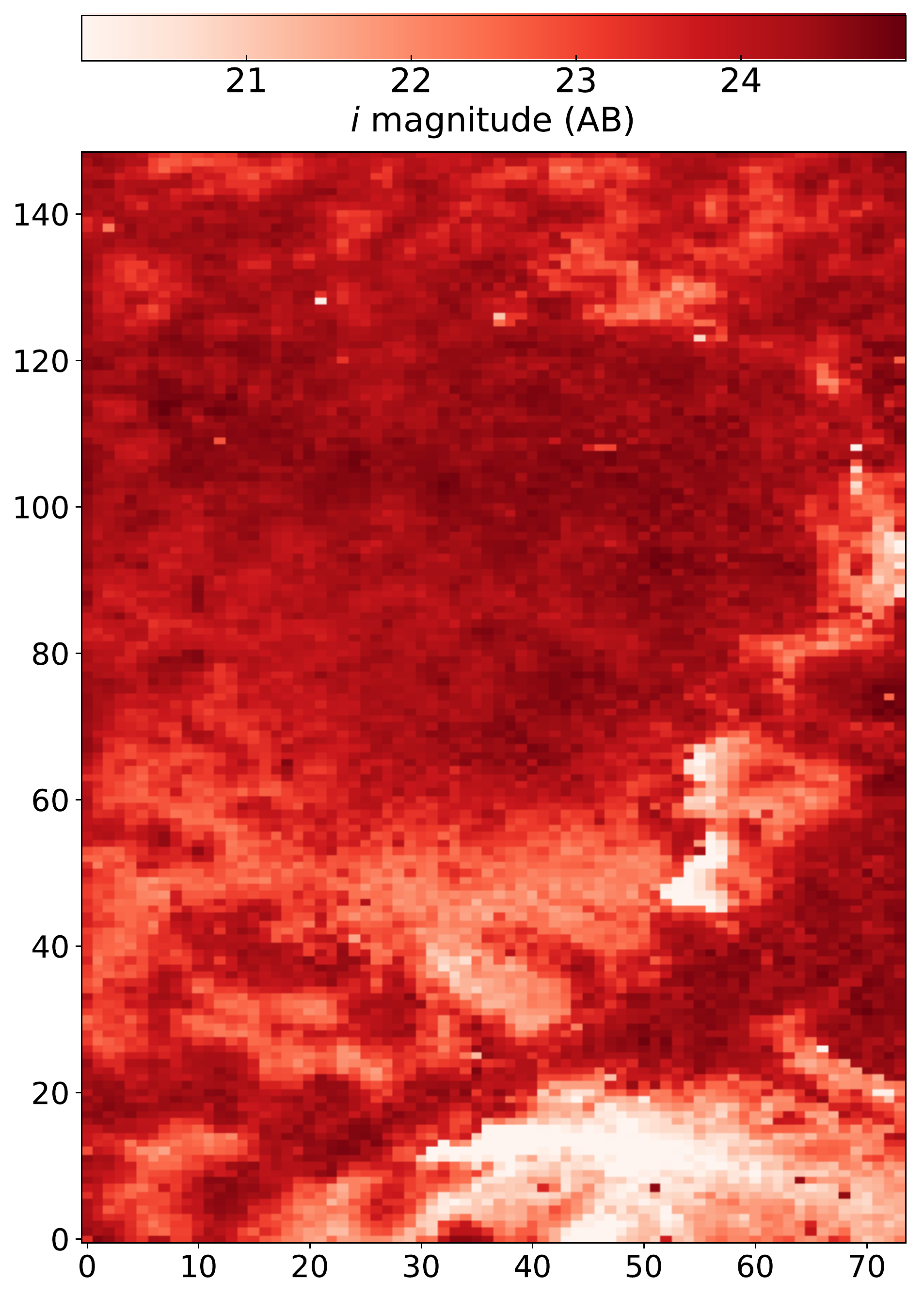} &
  	\includegraphics[width=0.32\textwidth]{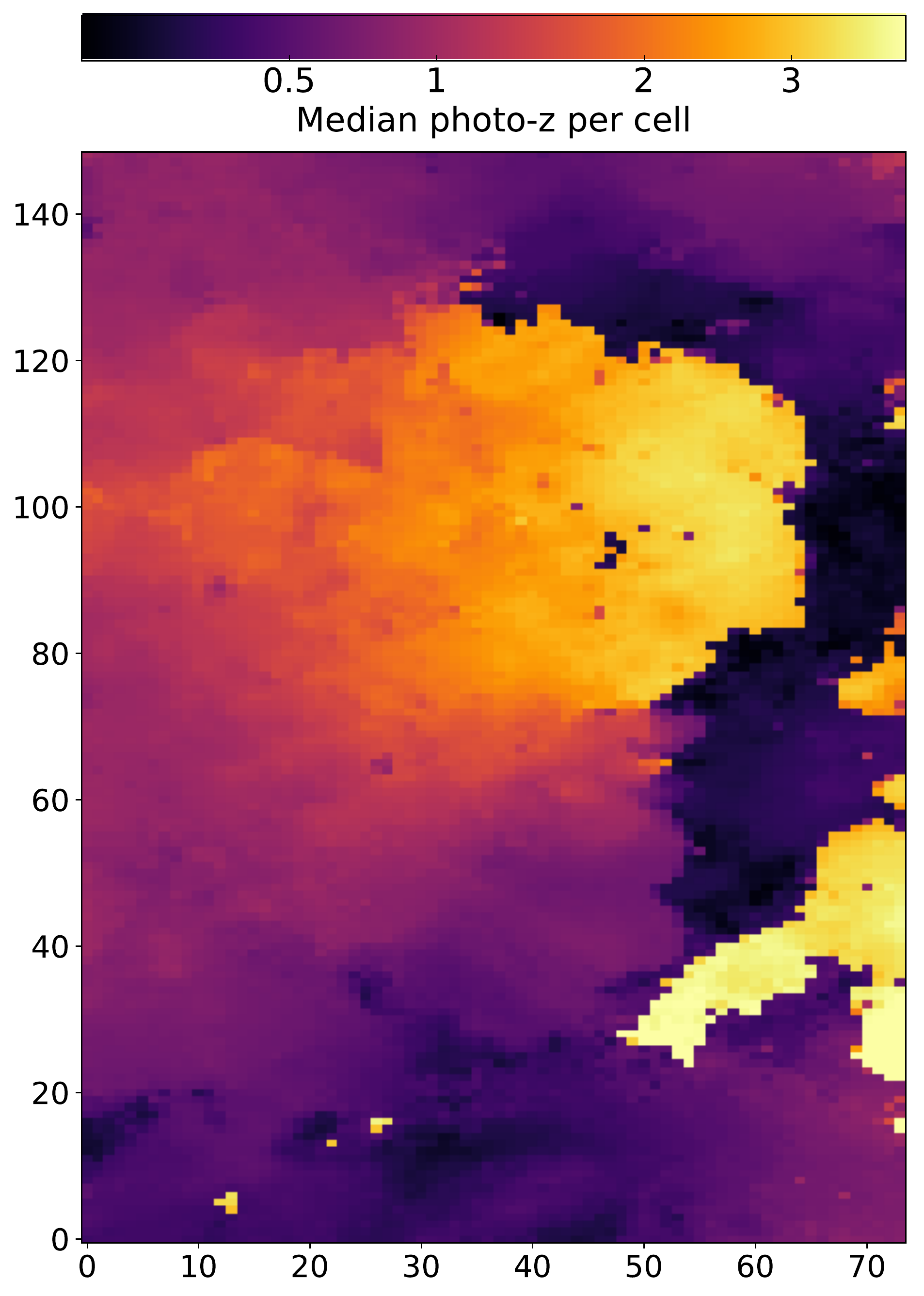} 
    \\    
  \end{tabular}
  \caption{Illustration of the updated self-organized map (SOM) used for the C3R2 DR2 source selection and analysis. Each cell of the SOM represents a particular  spectral energy distribution (SED) that shows up with regularity in the deep field data. The axes should be thought of as indices to parts of the high-dimensional galaxy color space. \emph{Left:} The map colored according to one of the features it encodes, namely the \emph{u}-\emph{g} color. \emph{Middle:} The map colored according to the median \emph{i} band magnitude of galaxies occupying each cell. It is clear that the typical magnitude is strongly color dependent. \emph{Right:} The SOM colored by the median photometric redshift of galaxies occupying each cell. Note the topological nature of the SOM: similar SEDs group together, producing relatively smooth variation of \phz\ with position on the map.}
      \label{figure:new_som}
\end{figure*}

\subsection{Validation of the photometry}
We examined star colors in a range of color-color diagrams to validate the photometry between fields. Stars were selected from each field using the cut in \textit{r} band magnitude  versus half-light radius shown in Figure~\ref{figure:cosmos_locus}. The example color-color plots shown in Figure~\ref{figure:stellar_locus} illustrate the color uniformity across broad wavelength baselines. The stellar loci align well, with residual offsets at the $\lesssim$0.02~magnitude level, which is sub-resolution of the SOM used for targeting. Moreover, we examined the spread of galaxy colors in each field with the well-known $gzK_{s}$ diagram (Figure~\ref{figure:gzK}). Again we see good agreement between fields. We therefore have confidence that galaxies in each field can be compared on the same color system.

\section{Revised target selection procedure}
In general, it is not possible to populate spectroscopic slit masks (covering $\sim$50~arcmin$^2$ of sky) exclusively with galaxies that are both important for photo-$z$ calibration and will yield secure redshifts with a given observational setup. We therefore developed a priority scheme that weights galaxies first according to their usefulness for photo-$z$ calibration, and then sorts based on the probability of obtaining a secure redshift for a planned observation, to maximize survey efficiency. The selection is ultimately based on the revised color map, which we now describe.

\subsection{Updated self-organized map}

We regenerated a self-organizing map to use as the basis for C3R2 targeting and analysis from the homogenized deep field photometry from VVDS-2h, COSMOS, and EGS. Projections of the new SOM are shown in Figure~\ref{figure:new_som}. The structure is similar (by design) to the SOM presented in M15 and used in M17. To generate this SOM, we initialized the algorithm with weight vectors from the previous SOM, and tuned down the responsiveness of the algorithm (learning rate function and neighborhood function) in order to preserve the previous structure as much as possible, while faithfully representing the data in the updated color system. For this SOM we included the $K_{s}$ filter. After the SOM was generated, we tested it by measuring the quantization error -- i.e., the typical color offset of a given galaxy from its best-matching cell. As in M15, we find an average Euclidean distance of a galaxy from its best-fit cell in the map of $\sim$0.2 mag, meaning that the individual colors of galaxies in the sample are typically offset from the colors of their best-matching cell in the SOM by $\lesssim$0.07 mag.

\subsection{Revised target weighting scheme}
 We begin with catalogs of uniform photometry in VVDS-2h, COSMOS and EGS, along with all available spectroscopic redshifts from these fields. Each galaxy in the catalogs was then matched to its best-fit SOM cell based on its photometry $-$  i.e., it was assigned to the SOM cell whose weight vector its colors most closely resemble. This sorting of galaxies on the SOM by color lets us develop a picture of where (in galaxy multi-color space) we have spectroscopy and where it is lacking.

We then cycle through each SOM cell to set priorities for sources in that cell. First we identify existing spectroscopic redshifts in the color cell, as well as existing but currently unanalyzed C3R2 spectra for sources in the cell. Sources with a previous redshift measurement or an unanalyzed C3R2 spectrum are given priority of zero. The algorithm used to set priorities for the rest of the sources is then as follows: 

\begin{enumerate}
\item{Set the source's initial priority based on the level of existing spectroscopic coverage for its color cell, increasing the priority as the quality of spectroscopic coverage diminishes:
\vspace{0.4pt}
\begin{itemize}
\item{If there is at least one high-confidence redshift in the cell already: initial weight = 1.}
\item{Low/medium-quality redshift and an unanalyzed C3R2 spectrum: initial weight = 2.}  
\item{Low/medium-quality redshift in cell: initial weight = 5.}
\item{Unanalyzed C3R2 spectrum in cell: initial weight = 7.}
\item{Neither existing redshifts nor unanalyzed C3R2 spectra in cell: initial weight = 15.} 
\end{itemize}}
\item{Divide weight by two if the object is a poor match to the colors represented by the cell. Specifically, do this if the Euclidean distance between the colors of the object and the cell is greater than 0.3 mag ($\sim$30\% of sources are color outliers at this level, corresponding to an average color difference of $\sim$0.1~mag in individual colors.) This step is done to avoid calibrating with sources with ``unrepresentative" colors.}
\item{Penalize objects missing two or more bands of photometry by dividing the weight by the number of missing bands.}
\item{Boost for color cell occupation: First quartile (those color cells with the lowest number of galaxies) multiply object weight by 100, second quartile multiply by 150, third quartile multiply by 200, fourth quartile (those color cells with the highest number of galaxies) multiply by 250.}
\item{Further down-weight sources in cells with \emph{multiple} reliable redshifts by dividing the weight by the number of reliable redshifts in the color cell.}
\end{enumerate}

The criteria described above are chosen to focus effort on well-populated color cells with little or no existing spectroscopy. Finally, once this priority scheme is applied, we break the catalog up into sources which can be expected to yield redshifts with different instrument/exposure times. For example, for a planned one-hour MOSFIRE \textit{H} band observation, we consider sources with expected required integration time $<$2h to be the top tier, and put them on the mask first. Sources predicted to take longer would be included in a second tier of the mask design. The predicted required exposure times are estimated with an exposure time calculator (written by P. Capak) taking into account the modeled detailed spectral characteristics of each source as well as the instrument sensitivity. For each galaxy we have a predicted type (star-forming, quiescent, or somewhere in between) and redshift from SED fitting. We can thus predict the strongest spectral feature that should be present for a given spectroscopic setup, and set a reasonable continuum S/N threshold for the feature to yield a firm redshift\footnote{The exposure time estimation is explained in more detail in the Appendix of M15.}. A variety of mask exposure times (from 1-6 hour) were employed to ensure that we would be sensitive to the range of sources in the high-priority sample.

\begin{deluxetable}{lcc}
\tabletypesize{\normalsize}
\tablecaption{Instruments and wavelength coverage \label{tab:coverage}}
\tablewidth{0pt}
\tablehead{
\colhead{Instrument} & \colhead{Wavelength range\tnote{\textdagger}\textdagger} & \colhead{Multiplex} \\
\colhead{} & \colhead{($\mu$m)} & \colhead{}
}

\startdata
DEIMOS & 0.5$-$1.0 & $\sim$90 \\
LRIS & 0.32$-$1.0 & $\sim$25  \\ 
MOSFIRE $H$ & 1.45$-$1.8 & $\sim$25  \\ 
\enddata
\tablecaption{}
\begin{tablenotes}
    \item[\textdagger] \textdagger Coverage varies somewhat from object to object. These numbers are typical.
    \end{tablenotes}
\end{deluxetable}

\section{Observations and Data Reductions}
The observations were conducted from September 2016 to April 2017. The observing nights and weather conditions are summarized in Table~2. Table~1 in the Appendix summarizes the observed slit masks.

\begin{figure*}[hbt]
\centering
  \includegraphics[width=\textwidth]{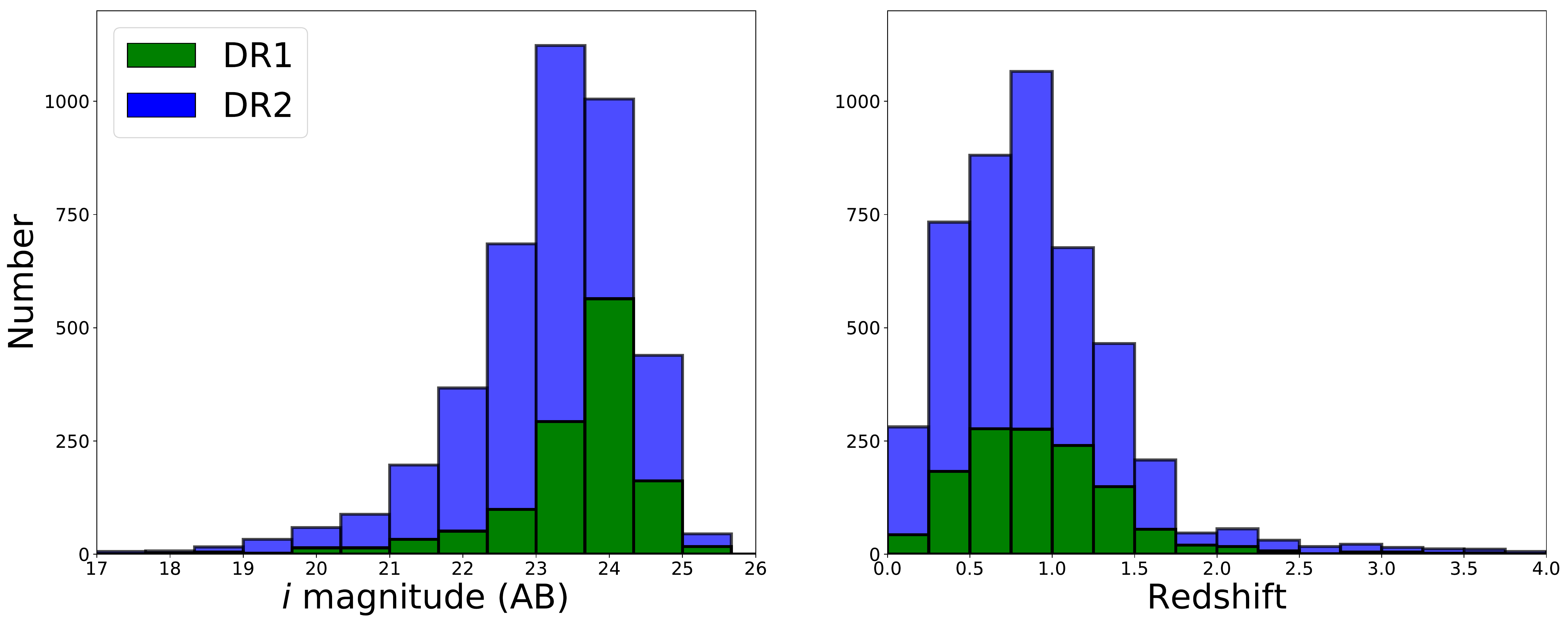} 
  \caption{Magnitude and redshift distributions for the C3R2 spectroscopic survey.}
      \label{figure:hists}
\end{figure*}

\subsection{Description of Observations}

Observations in 2016B/2017A were conducted largely as described in M17 for the DR1 data. Here we give a brief overview of the observations and reductions.

\subsubsection{DEIMOS}
DEIMOS observations were conducted using the 600~groove~mm$^{-1}$ grating blazed at 7200~\AA\ and the GG400 blocking filter, with dithering performed to improve  sky subtraction. We settled on a minimum slit length of 8$''$ as a balance between maximizing the number of targets on the mask and getting good sky measurements. Data were reduced using a modified version of the DEEP2 pipeline  designed to deal with dithered data.

\subsubsection{MOSFIRE}

MOSFIRE was used in its default configuration. For instrumental details we refer the reader to \citet{Steidel14}. For our \textit{H} band observations we used integration times of 120s with ABAB dithering to improve sky subtraction. Reductions were performed with the MOSFIRE Data Reduction Pipeline (DRP) made available by the instrument team\footnote{https://keck-datareductionpipelines.github.io/MosfireDRP/}. We chose to observe only in \textit{H} band because we discovered that the density of high-priority targets for which we could expect to get a secure redshift was notably higher in this band.

\subsubsection{LRIS}

We used the 400 groove~mm$^{-1}$ blue grism blazed at 3400~\AA\ and the 400 groove~mm$^{-1}$ red grating blazed at 8500~\AA, with the D560 dichroic. Our choice of blue grism gives high sensitivity at bluer wavelengths where spectral features are likely to be found for objects with photometric redshifts of $z\sim1.5-3$, while the red coverage allows for the detection of \oii\ for some sources out to $z\sim1.6$. The LRIS spectra were reduced using the IRAF-based BOGUS software developed by D. Stern, S. A. Stanford, and A. Bunker, and flux calibrated using observations of standard stars from  \citet{Massey90} observed on the same night using the same instrument configuration.

\begin{deluxetable*}{lllcll}
\tablecaption{List of observing nights.}
\tablefontsize{\footnotesize}
\tablehead{
\colhead{UT Date} &
\colhead{Code} &
\colhead{Instrument} &
\colhead{\# Masks} &
\colhead{Observers} &
\colhead{Notes}}
\startdata
2016 Sep 29 & N06-L & LRIS    & 3 & AP,AS,DS       & 0\farcs6 seeing; not photometric \\
2016 Oct 02 & N07-D & DEIMOS  & 1 & DM,AP          & moisture/cloud cover, 0\farcs9 seeing when open \\
2016 Oct 03 & N08-D & DEIMOS  & 3 & DM,AP          & moisture/cloud cover,0\farcs8 seeing when open \\
2016 Nov 27 & N09-L & LRIS    & 1 & PB,MH,ARi,DS    & cloudy, variable seeing (1\farcs2 - 2\farcs0) \\
2016 Nov 28 & N10-D & DEIMOS  & 6.5 & AS,DS        & cloudy, variable seeing (1\farcs0 - 2\farcs0) \\
2016 Nov 29 & N11-D & DEIMOS  & 5 & AS,DS          & clear, 0\farcs65 seeing \\
2016 Nov 30 & N12-D & DEIMOS  & 2.5 & AS,DS        & intermittent closures due to fog, flurries; 1\farcs2 seeing \\
2016 Dec 01 & N13-D & DEIMOS  & \nodata & AS,DS    & lost to snow \\
2016 Dec 07 & N14-D & DEIMOS  & \nodata & PB,AG,BS,DS & lost to recovery from winter storm \\
2016 Dec 08 & N15-D & DEIMOS  & 1 & AG,MJ,BS,DS    & high humidity, cirrus; closed most of night \\
2016 Dec 20 & N16-D & DEIMOS  & \nodata & DS & [1/2 night]; lost to ice-covered dome\\
2017 Jan 02 & N17-D & DEIMOS  & 5 & DS & photometric; 0\farcs6 - 1\farcs0 seeing \\
2017 Jan 05 & N18-D & DEIMOS  & 5 & JC,NH,AS       & photometric, windy, poor seeing ($>$1\arcsec) \\
2017 Jan 06 & N19-D & DEIMOS  & 3 & JC,NH,AS       & 1/2 night lost to high humidity \\
2017 Jan 27 & N20-D & DEIMOS  & 4.5 & ARe,DS,AW    & photometric, 1\farcs0 seeing \\
2017 Jan 27 & N21-L & LRIS    & 5 & ARe,DS,AW      & $\prime\prime$ \\
2017 Jan 30 & N22-D & DEIMOS  & 4.5 & ARe,AW       & photometric, 0\farcs75 seeing\\
2017 Mar 31 & N23-D & DEIMOS  & 1 & JCh,JCo,NH     & poor, high humidity, intermittent closures \\
2017 Apr 01 & N24-D & DEIMOS  & 3.5 & JCo,NH       & photometric, 0\farcs9 seeing \\
2017 Apr 01 & N25-L & LRIS    & 2 & JCh,DS           & $\prime\prime$ \\
2017 Apr 02 & N26-D & DEIMOS  & 2.5 & JCo,NH       & photometric, 1\farcs0 seeing \\
2017 Apr 02 & N27-L & LRIS    & 2 & JCh              & $\prime\prime$ \\
2017 Apr 15 & N28-M & MOSFIRE & 3 & NH,DM,DS       & photometric, 0\farcs7 seeing\\
2017 Apr 16 & N29-M & MOSFIRE & 4 & NH,DM,DS       & photometric, 0\farcs6 seeing\\
\enddata
\label{table:observations}
\tablecomments{Observers (alphabetical by last name):  
PB -- Peter Boorman,
AG -- Audrey Galametz,
JCh -- Jason Chu,
JCo -- Judith Cohen,
MH -- Marianne Heida,
NH -- Nina Hernitschek,
MJ -- Marziye Jafariyazani,
DM -- Daniel Masters, 
BM -- Bahram Mobasher,
AP -- Andreas Plazas,
ARe -- Alessandro Rettura,
ARi -- Adric Riedel,
BS -- Behnam Darvish Sarvestani,
AS -- Adam Stanford,
DS -- Daniel Stern,
AW -- Anna Weigel.}
\end{deluxetable*}

\subsection{Redshift determination}

We refer the reader to M17 for a detailed description of the redshift determination procedure, as well as the quality flags and failure codes we adopt. Briefly, each observed source was assessed independently by two co-authors to determine the redshift and associated quality flag ($Q=0-4$, with 4 being certainty), and any conflicts were reconciled through a joint review of the spectra with the help of a third, independent reviewer.  Sources for which we failed to identify a redshift were assigned failure codes to indicate the most likely reason: 
\begin{itemize}
    \item{Code = $-$91: Insufficient S/N;}
    \item{Code = $-$92: Well-detected but with no discernible features;}
    \item{Code = $-$93: Problem with the reduction;}
    \item{Code = $-$94: Missing slit (essentially an extreme case of $-$93). }
\end{itemize}

As in M17, we also investigated all Q=4 (highest quality) sources for which the spectroscopic redshift ($z_s$) was highly discrepant from the expected photometric redshift.  Specifically, we investigated all sources with $\lvert z_{p} - z_{s} \rvert / (1 + z_{s}) \geq 0.15$. The results are presented in \S5.4.

\section{Redshift Results}

Out of 4407 targeted sources in DR2, we find a high-confidence ($Q$~$\geq$~3) redshift for 2809, for an overall success rate of 64\%. If we consider only the photometric observing nights, our success rate is $\sim$70\%. We also obtained redshifts for 433 serendipitous sources, for a total of 3242 high-quality redshifts in this release. Some of these (71) are unintentional duplicate observations (see \S5.3), therefore the total number of unique redshifts released is here is 3171. Combined with the 1283 sources published in M17, the survey to date has obtained and published 4454 redshifts. The magnitude and redshift distributions of the C3R2 sample are given in Figure~\ref{figure:hists}. The C3R2 redshift catalog is provided in a machine-readable table, the contents of which are summarized in Table~2 of the Appendix. 

\subsection{Redshift performance of the SOM technique}

As a first-order test of the SOM-based calibration method, we assign all galaxies in our spectroscopic sample (DR1+DR2) a \phz\ estimate based entirely on their position on the SOM. First, every cell in the SOM is assigned a ``calibrated'' redshift based on the median \phz\ of all galaxies occupying that cell, where the \phz s come from the literature \citep{Ilbert06, Laigle16}. This calibrated SOM map is illustrated in the right panel of Figure~\ref{figure:new_som}. 

Galaxies are then assigned a SOM-based redshift point estimate (a ``SOMz") according to which SOM cell they belong to. Note that no spectroscopic information is directly used in calibrating the mgap at this stage; the calibration is based on averaging \phz\ estimates  (which may come from data with more bands / higher dimensionality, as with the COSMOS 30-band \phz\ estimates) at each point in the \e/\w\ color space.

\begin{figure*}[htb]
\centering
    \includegraphics[width=0.98\textwidth]{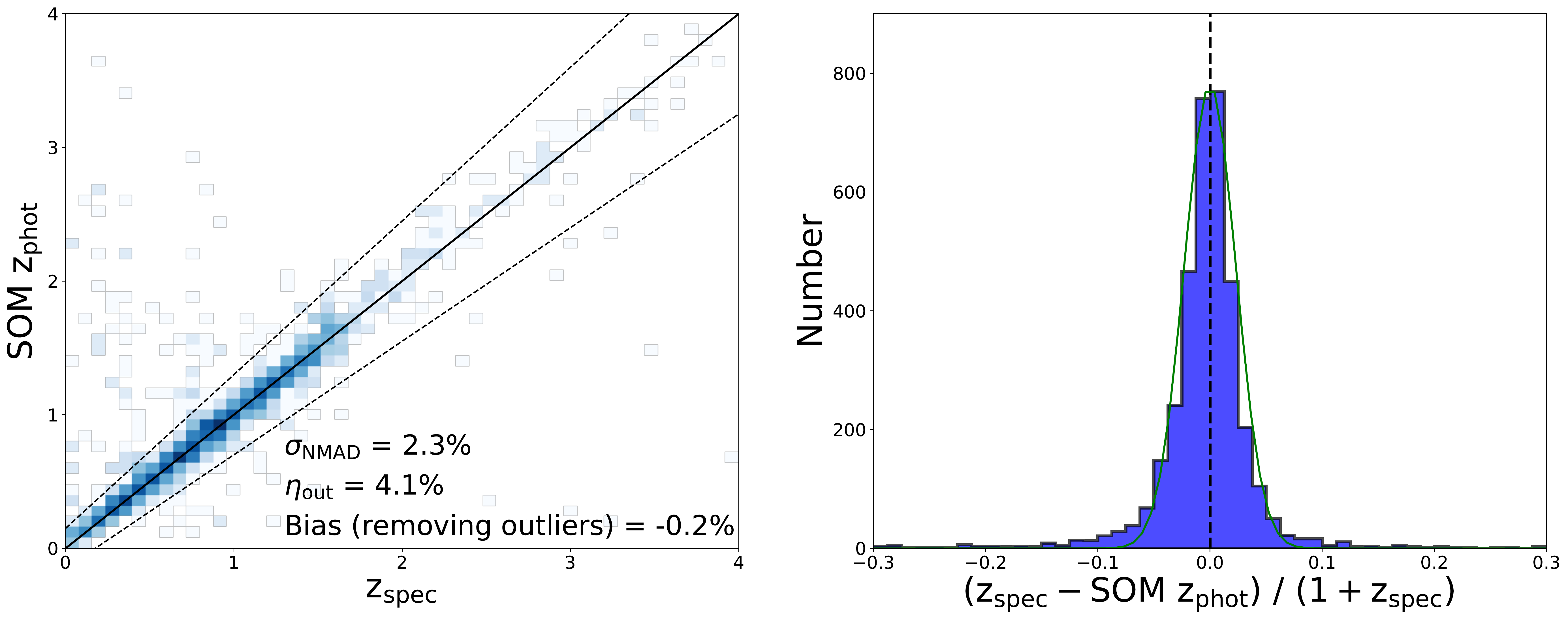} 
  \caption{The performance of the SOM method on the DR1+2 data. We find an outlier fraction of 4.1\%, a scatter of 2.3\%, and a bias after removing outliers of -0.2\%. Also, it is worth noting that our outliers are mostly low redshift sources which scattered to high SOM\textit{z} rather than the reverse, which may be an observational bias as we are unable to measure redshifts for many of the outliers on the other side, with true redshifts at $z>1.5$. The green line on the right panel is a Gaussian centered at 0 with $\sigma=2.3$\%.}
      \label{figure:performance}
\end{figure*}

With our ``gold'' sample ($Q=4$) redshifts from DR1+2 we get the results shown in Figure~\ref{figure:performance}. The outlier fraction, defined as those sources with $\lvert z_{p} - z_{s} \rvert / (1 + z_{s}) > 0.15$, is 4.1\%. The scatter estimated using the normalized median absolute deviation ($\sigma_\mathrm{NMAD}$) is 2.3\%. The overall bias of the sample (after removing outliers) is $-$0.2\%. 

For comparison, we gave the same photometry used to infer SOM-based \phz s to the template fitting code \textit{Le Phare} \citep{Arnouts99, Ilbert06}. We find a scatter $\sigma_\mathrm{NMAD}$ of 2.9\%, an outlier fraction of 3.8\%, and bias after removing outliers of $-1.3$\%. The scatter is significantly higher than with the SOM-based \phz, while the outlier rate is slightly lower. Importantly, the bias is substantially worse. This illustrates the performance gain achieved simply by averaging high-quality \phz\ estimates (derived from overlapping deep data, potentially with additional bands) at each point in the lower-dimensional color space of a wide-area survey such as \e\ or \w.

\subsection{Serendipitous sources}

We detected 437 sources that serendipitously landed in a slit with the primary source (``serendips"). Of these, 51 were already in the literature with high-confidence redshifts. The agreement with literature values is nearly exact aside from three sources. We believe we understand these mismatches and trust our results in those cases. 

A number of our serendipitous detections fall very close ($\lesssim$1$\arcsec$) to the main source of interest, highlighting an important issue for photometric redshift calibration: the possibility of obtaining a secure redshift for the wrong source. In fact, in some of these cases it is difficult or impossible to determine which source is the primary and which is the serendipitous source. This issue was also noted by \citet{Brinchmann17}, who find that up to a few percent of secure emission line redshifts in deep surveys can be spurious, due to a background source blended with the target of interest. 

\begin{figure*}[htb]
\centering
    \includegraphics[width=0.95\textwidth]{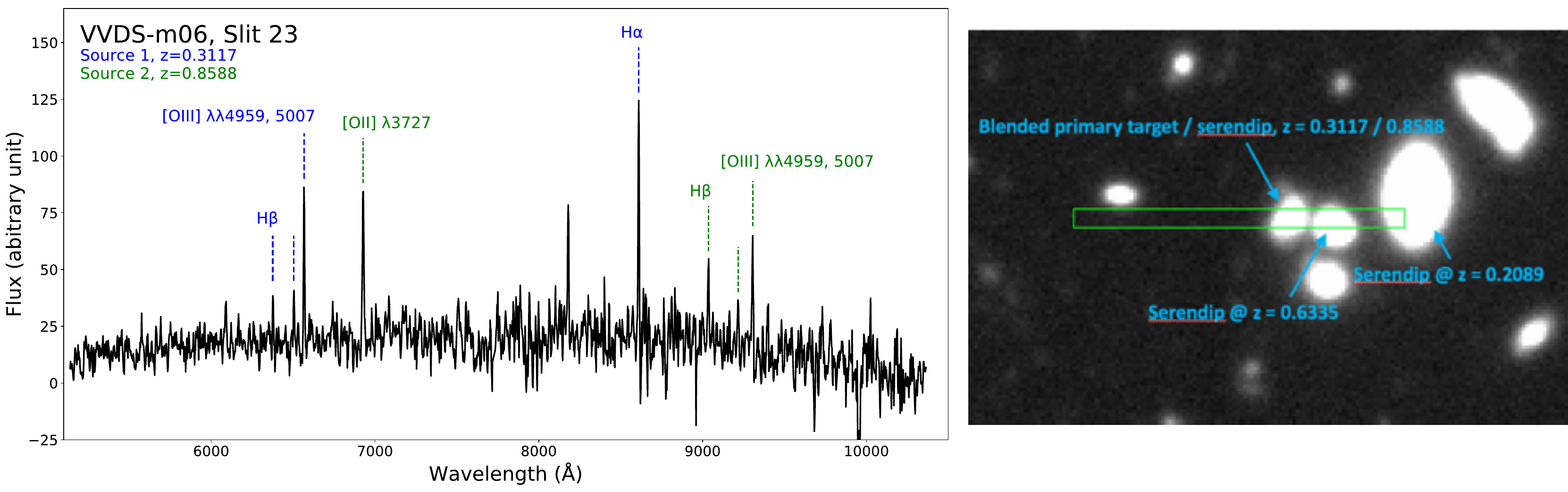}
  \caption{An example in which we clearly detect two emission line redshifts for what appears to be a single source. This slit is particularly busy as there are an additional two serendipitous sources at $z=0.6335$ and $z=0.2089$ (indicated on the CFHTLS image on the right, with DEIMOS 1$^{\prime\prime}$ width slit overlaid) which were detected; these spectra are not shown in the 1-D extraction on the left. In the spectrum shown,  the two redshifts reveal the problematic nature of the source. However, it is easy to imagine more pathological cases for which there is only one emission line redshift detected and incorrectly attributed to the primary target. In any case, as there is clearly blended photometry for this source and similar ones should likely be excluded from calibration samples.}
      \label{figure:overlap}
\end{figure*}

In any case it is clear that such sources will have blended photometry, and should likely be considered invalid for redshift calibration purposes. We summarize the close serendips in Table~3, and illustrate one example in which two sources completely overlap in Figure~\ref{figure:overlap}. In that case two redshifts are clearly evident, but one could imagine cases in which a single secure, but spurious, emission line redshift is obtained for the primary source due to an overlapping galaxy.

\subsection{Duplicate observations}
We (unintentionally) obtained duplicate high-confidence redshifts for 71 sources, providing a useful check on the consistency and precision of our redshift results. We found no inconsistent redshift results, and our redshifts are precise to $\lesssim 1\times10^{-4}$; the median value of $\Delta z /(1+\bar{z})$ between duplicate \sz s is $1.1\times10^{-4}$. Note that the duplicate results are published in the catalog, while the numbers quoted for the overall survey are for the unique sources in the sample.

\begin{deluxetable*}{lclccc}
\tablecaption{Close serendipitous detections, falling within $1\arcsec$ of the primary target.}
\label{tab:close_sers}
\tablewidth{0pt}
\tablefontsize{\footnotesize}
\tablehead{
\colhead{Mask} & \colhead{Slit} & \colhead{Primary target name} & \colhead{Primary \sz} & \colhead{Serendip \sz} &\colhead{Offset} }

\startdata
VVDS-m01 &      83 & VVDS-45179 & 1.2477 & 1.2480 & 0.00$\arcsec$ \\ 
EGS-H-2hr3 &       6 & EGS-336001 & 1.5624 & 1.7120 & 0.00$\arcsec$ \\ 
VVDS-m22 &      67 & VVDS-186226 & 0.4635 & 1.4900 & 0.00$\arcsec$ \\ 
VVDS-m06 &      23 & VVDS-439390 & 0.3117 & 0.8588 & 0.00$\arcsec$ \\ 
EGS-4h1 &      58 & EGS-312294 & 2.8651 & 1.5250 & 0.00$\arcsec$ \\ 
EGS-4h1 &       4 & EGS-305620 & 1.4168 & 0.7322 & 0.01$\arcsec$ \\ 
EGS-4h1 &      25 & EGS-298760 & 0.4723 & 0.8775 & 0.01$\arcsec$ \\ 
COSMOS-5h3 &      64 & COSMOS-296536 & 0.9890 & 1.1605 & 0.01$\arcsec$ \\ 
COSMOS-5h3 &       5 & COSMOS-293140 & 0.7493 & 0.9665 & 0.01$\arcsec$ \\ 
COSMOS-6h &      87 & COS-100 & 1.5746 & 0.7375 & 0.02$\arcsec$ \\ 
COSMOS-1h2 &      78 & COSMOS-166492 & 1.3604 & 1.4165 & 0.02$\arcsec$ \\ 
COSMOS-m22 &      61 & COSMOS-263650 & 1.3413 & 0.9970 & 0.12$\arcsec$ \\ 
COSMOS-6h &      38 & COS-980010 & 1.6202 & 0.8337 & 0.22$\arcsec$ \\ 
EGS-H-2hr5 &       4 & EGS-394269 & 1.5812 & 1.3597 & 0.25$\arcsec$ \\ 
COSMOS-6h &      85 & COS-999227 & 1.2182 & 0.8367 & 0.29$\arcsec$ \\ 
COSMOS-m12 &      27 & COSMOS-490417 & 1.4051 & 0.4745 & 0.35$\arcsec$ \\ 
EGS-4h2 &      17 & EGS-372214 & 1.2711 & 1.1722 & 0.38$\arcsec$ \\ 
COSMOS-m21 &      78 & COSMOS-238677 & 1.1551 & 1.2177 & 0.41$\arcsec$ \\ 
EGS-4h2 &      43 & EGS-353596 & 1.6282 & 0.7286 & 0.44$\arcsec$ \\ 
VVDS-m25 &      18 & VVDS-271560 & 0.3728 & 1.5569 & 0.55$\arcsec$ \\ 
VVDS-m02 &      68 & VVDS-94232 & 0.9616 & 0.7872 & 0.63$\arcsec$ \\ 
VVDS-m04 &      75 & VVDS-261586 & 0.6941 & 0.6346 & 0.66$\arcsec$ \\ 
COSMOS-m28 &      53 & COSMOS-460081 & 0.8918 & 0.8332 & 0.68$\arcsec$ \\ 
VVDS-m14 &      46 & VVDS-144620 & 0.6213 & 0.6215 & 0.70$\arcsec$ \\ 
COSMOS-m24 &      82 & COSMOS-324794 & 0.7317 & 0.7253 & 0.72$\arcsec$ \\ 
EGS-4h1 &      42 & EGS-317641 & 0.8442 & 0.9774 & 0.72$\arcsec$ \\ 
COSMOS-m17 &      39 & COSMOS-105578 & 1.1781 & 0.4723 & 0.73$\arcsec$ \\ 
EGS-4h2 &      86 & EGS-359810 & 0.6786 & 1.2090 & 0.74$\arcsec$ \\ 
VVDS-m12 &      66 & VVDS-215721 & 1.1173 & 0.8511 & 0.74$\arcsec$ \\ 
EGS-4h2 &       9 & EGS-362393 & 1.3213 & 0.9630 & 0.75$\arcsec$ \\ 
VVDS-m01 &      29 & VVDS-56665 & 0.9600 & 0.9004 & 0.77$\arcsec$ \\ 
COSMOS-m33 &      22 & COSMOS-965712 & 0.2236 & 1.4478 & 0.83$\arcsec$ \\ 
COSMOS-m28 &      71 & COSMOS-451219 & 0.1013 & 1.2030 & 0.83$\arcsec$ \\ 
VVDS-m05 &      41 & VVDS-344788 & 0.4372 & 0.6962 & 0.91$\arcsec$ \\ 
EGS-4h2 &       7 & EGS-373437 & 1.7616 & 0.3345 & 0.91$\arcsec$ \\ 
VVDS-m07 &      65 & VVDS-499386 & 0.8773 & 0.3455 & 0.97$\arcsec$ \\ 
\enddata

\end{deluxetable*}

\begin{figure*}[htb]
\centering
  \begin{tabular}{@{}ccc@{}}
      \includegraphics[width=.3\textwidth]{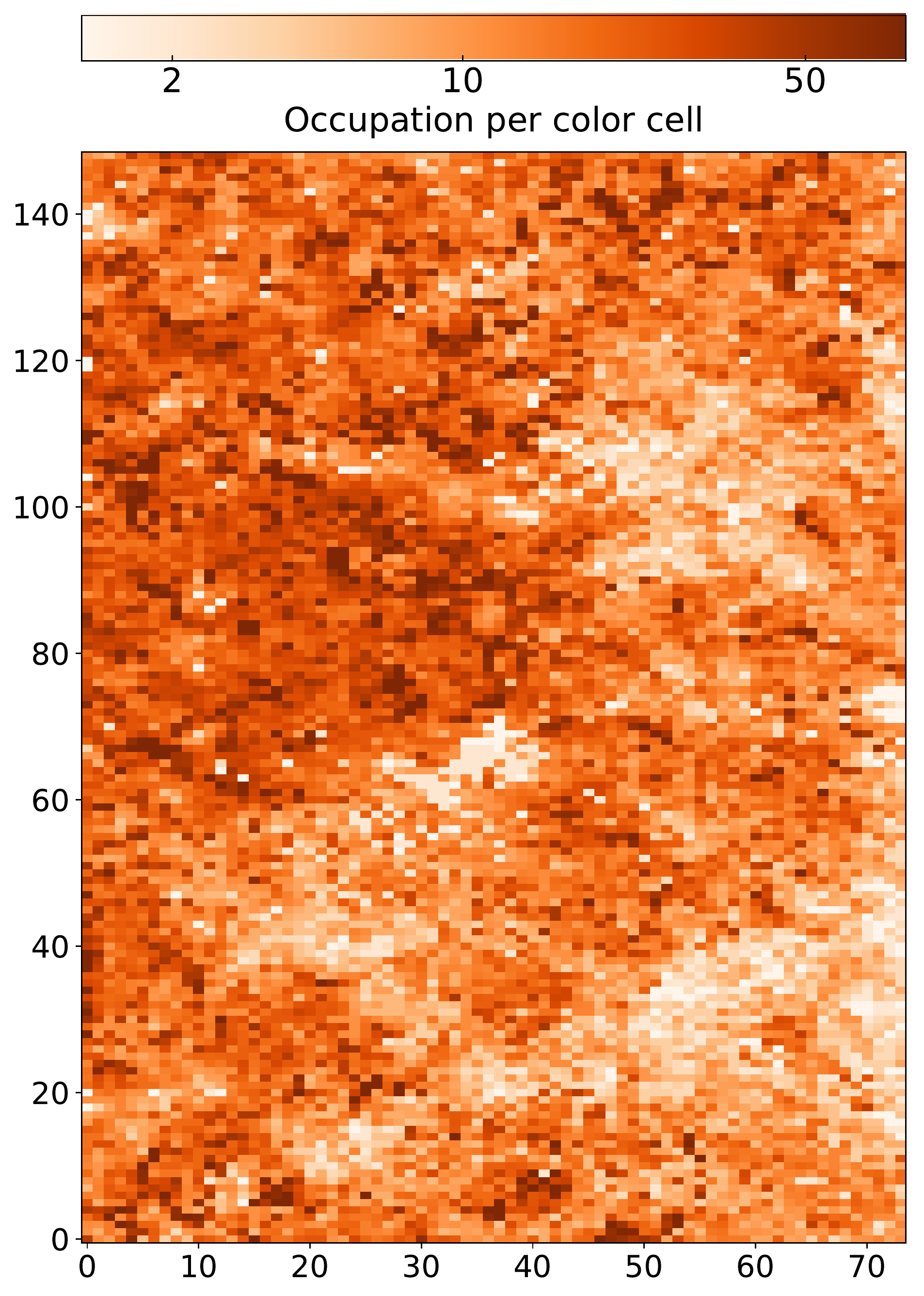} &
    \includegraphics[width=.3\textwidth]{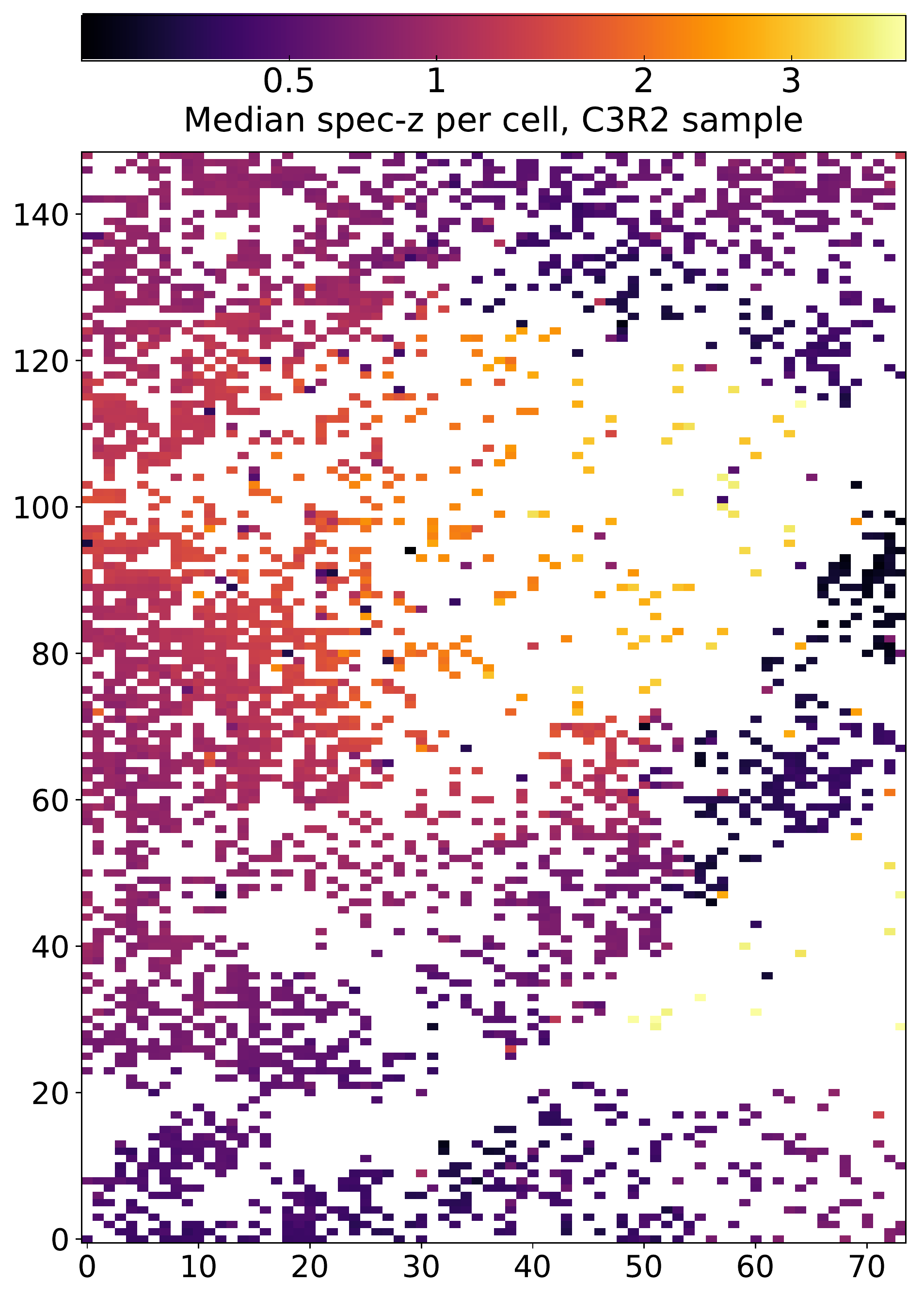} &
  	\includegraphics[width=0.3\textwidth]{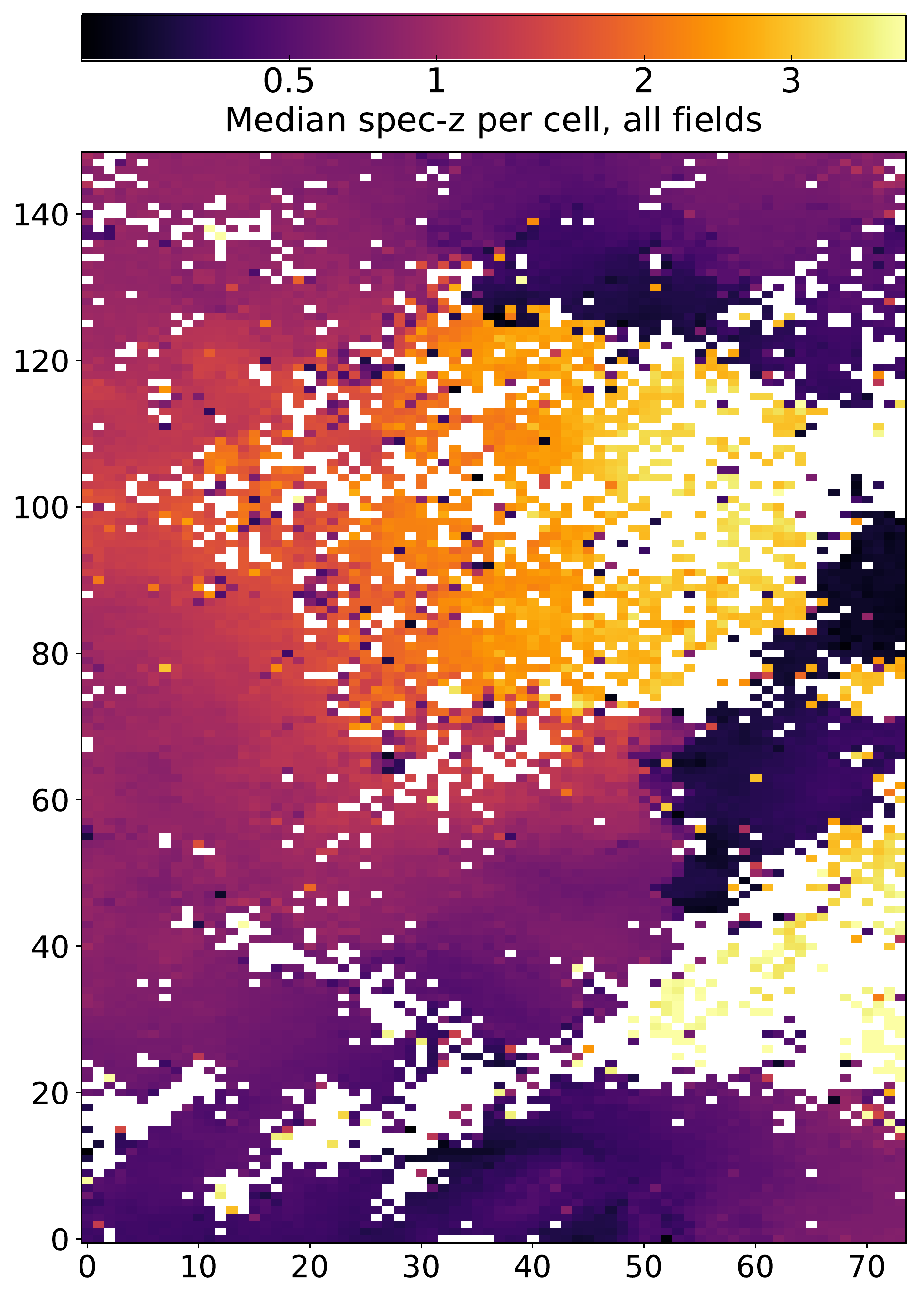}
    \\    
  \end{tabular}
  \caption{\textit{Left}: Occupation density of galaxies on the SOM. The distribution is clearly not uniform, reflecting the different sky density of galaxies in different parts of the \e/\w\ color space. \textit{Center}: C3R2 coverage of the color space. C3R2 has successfully observed $>$35\% of the color cells. \textit{Right}: The current total coverage of the SOM, incorporating C3R2 with spectra from numerous other surveys. Roughly 76\% of color cells are covered, while $\sim$85\% of galaxies live within calibrated cells. Comparison with the occupation plot on the left shows that the remaining cells strongly correlate with less occupied regions of the galaxy color space.}
      \label{figure:coverage}
\end{figure*}

\subsection{Assessing the photo-z outliers}

We examined the 130 outliers, or those sources with a secure redshift that differed from the predicted SOM-based redshift by $>$15\%. These fall in a range of categories: $\sim$25\% have an individual \phz\ estimates more in line with the \sz, hinting at a true degeneracy in the color-redshift relation when limited to the \e/\w\ color space; $\sim$25\% can be understood as rare objects (low-mass stars, quasars, extremely strong line emitters) with correspondingly unusual colors; while another $\sim$20\% have unusual colors due to blending with a nearby galaxy. The remaining $\sim$30\% seem to be either true outliers for which both the SOM-\textit{z} and \phz\ are incorrect, or problems with our redshift assignment. In fact, we found 14 cases for which the assigned redshift or quality flag was incorrect, either due to a mistake in the original assessment of the spectrum or in the process of reconciling and cataloging the results. We modified these accordingly\footnote{One might worry that modifying redshifts ``by hand" after the fact could be injecting a bias in the results. However, we are concerned with producing the highest quality data product possible. We therefore feel it is justified if obvious mistakes are uncovered; moreover, the total number of cases amounts to  only 0.4\% of the overall sample.}. This analysis highlights the necessity of using great care in developing the ``gold sample" of galaxies to use in calibrating \phz\ for cosmology.


\subsection{Spectroscopic failures}

Unfortunately we cannot achieve 100\% redshift success for faint galaxies. We therefore made an attempt to understand our spectroscopic failures and potential biases in the data. We consider only spectra from good weather nights, defined as $\leq1\arcsec$ seeing with photometric conditions. We targeted a total of 4240 sources under these conditions in DR1+2. Of these, 1287 (30\%) were ``failures", meaning we could not recover a secure redshift from the reduced spectrum. 

We focus on the good-weather failures in the COSMOS field, for which we have the most accurate photometric redshifts, ancillary data, and predicted exposure times from our calculator. This leaves 671 sources. Of these, 186 were assigned low-confidence ($Q=2$ or $Q=1$) redshifts, 286 have a failure code of $-$91, indicating they were too faint to yield a secure redshift, and 149 have a code of $-$92, indicating a reasonable continuum detection but nevertheless no redshift determination. The remaining 50 have code of -93 or -94, meaning there was a reduction problem, so they are not of great interest for this analysis. 

Of the sources too faint to yield a secure redshift, 71\% were predicted by our exposure time calculator to take longer than the actual integration; in other words, we \textit{expected} to fail for these sources. For the other 29\% the situation is less clear, but it is likely that our exposure time calculator was optimistic. For the failures for which we had a good continuum detection, a full 66\% were not expected to yield a secure redshift for the given exposure time. Of the others, $\sim$38\% were classified as quiescent, indicating that we have systematic difficulty (as expected) getting redshifts for passive galaxies in comparison to star-forming ones.

A potentially worrying failure mode for calibration arises from sources at the redshift boundary of what can easily be done with DEIMOS or other optical spectrographs. Nearly 50\% of the sources for which we measured continuum but no redshift have a \phz\ near 1.5, where the \oii\ line can easily be missed if it happens to fall off the red end of the detector. One can imagine a bias toward lower redshifts when calibrating such galaxies.

\begin{figure*}[htb]
\centering
    \includegraphics[width=0.85\textwidth]{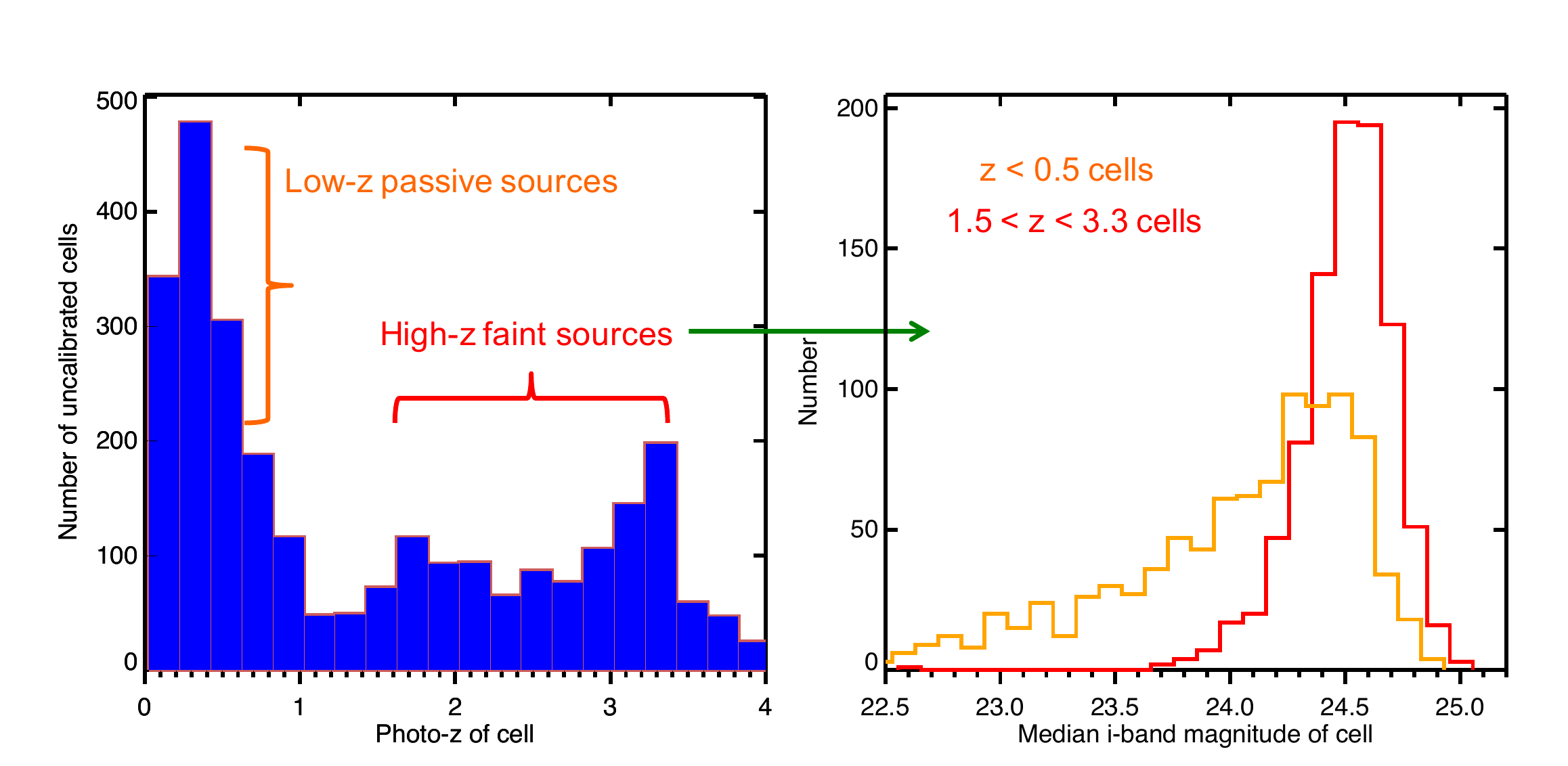} 
  \caption{Summary of the presently uncalibrated cells. Most of the cells correspond to relatively low-density parts of the color space. The low-\textit{z} peak represents mostly passive galaxies for which obtaining a secure redshift is difficult (as well as some star-forming but rare sources). The high-\textit{z} peak are galaxies for which obtaining a redshift with an optical spectrograph is difficult, and are primarily quite faint, as shown on the right-hand plot.}
      \label{figure:uncalibrated}
\end{figure*}

\begin{figure*}[htb]
\centering
 \includegraphics[width=0.6\textwidth]{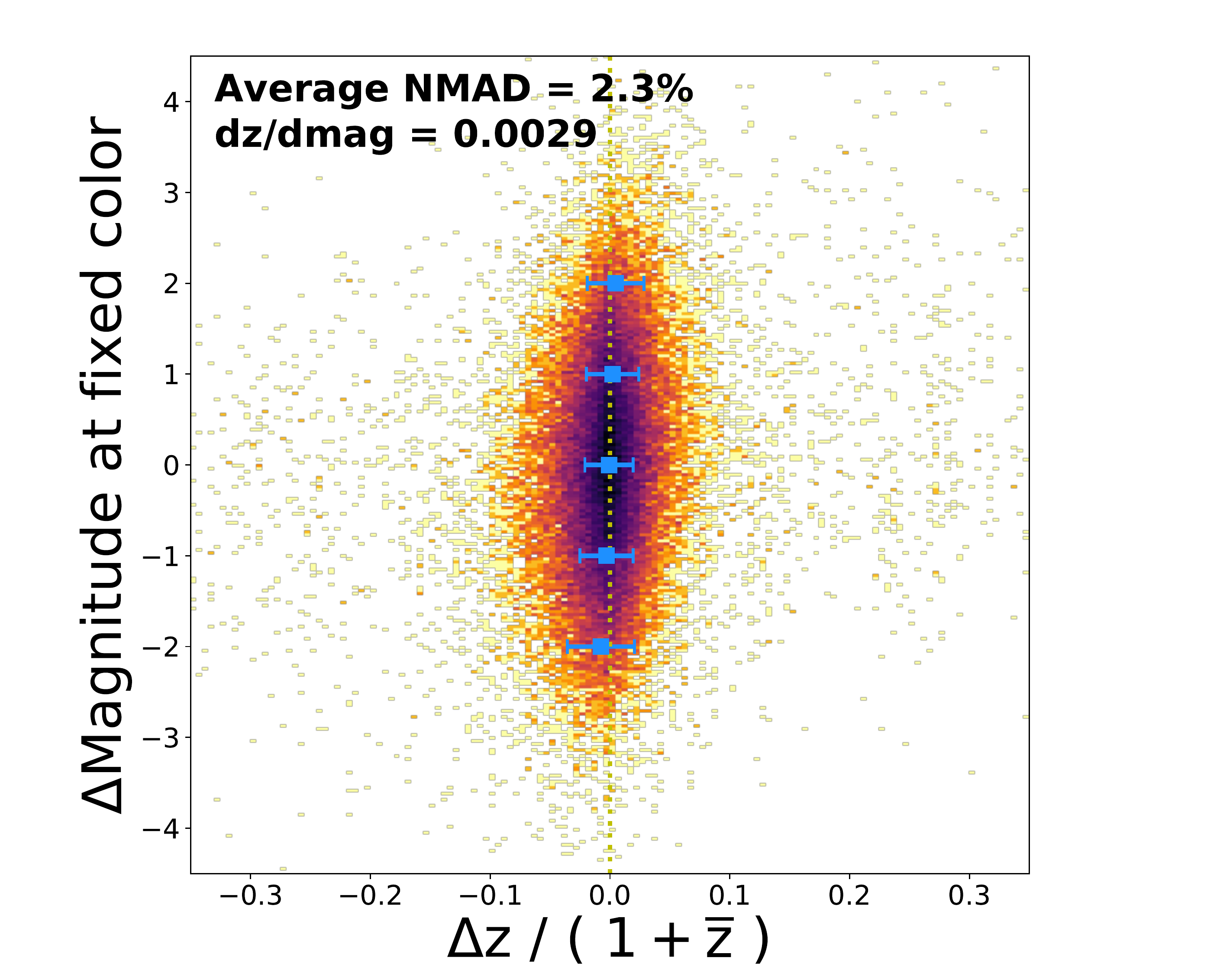} 
  \caption{A test of the importance of galaxy brightness for redshift inference for the \e/\w\ color space. For each unique pair of \sz\ galaxies at fixed color (in the same SOM cell), we plot the difference in $i$-band magnitude versus the difference in measured redshift. As can be seen, there is a weak dependence on magnitude, which we quantify with a fit to the running medians (in blue, with scatter bars, computed in slices of $\Delta$mag of width 0.2 across the distribution) as $\sim$0.0029 change in $\Delta z/(1+\bar{z})$ per magnitude. The scatter reflects the redshift uncertainty at fixed color, which depends on both the photometric error and intrinsic uncertainty in the color-redshift relation. This result has important implications for the \w\ calibration (see \citealp{Hemmati18}), for which there are significant numbers of optically faint sources with similar colors as brighter galaxies.}
      \label{figure:mag_prior}
\end{figure*}

\section{Calibration progress}

We performed some tests to get a sense of the state of redshift calibration. We will address the issue of large scale structure variations within the relatively small calibration fields in more detail in a separate paper. 

\subsection{Color space coverage}
When we consider spectra from all of our fields, we are currently covering $\sim$76\% of the cells with high-quality spectra, which ``calibrates'' $\sim$86\% of sources (Figure~\ref{figure:coverage}). The majority of the uncalibrated cells also are those with fewer galaxies, so in principle are less important to the overall calibration. For comparsion, prior to the C3R2 survey the color space coverage estimated in M15 was $\sim$50\%. The C3R2 sample alone has covered over 35\% of the color space, much of which was previously uncharted. 

It may be noted that the fraction of calibrated cells we quote may be dependent on the total number of cells, which is a hyperparameter of the SOM. We argue that the cells of the current SOM are fine enough that increasing the resolution further would not improve our understanding of the \pz\ relation at the depth of \e. When we say that a galaxy is ``calibrated" we are really saying that we have at least one spectroscopic redshift for a galaxy with colors that are indistinguishable from that galaxy at the depth of \e.

\subsection{What are the remaining uncalibrated galaxies?}

A key question for calibration is: What galaxies are we still missing, and why? The majority seem to be a combination of low-z quiescent or star-forming but rare  galaxies, and galaxies at $1.5\lesssim z \lesssim 3$ for which DEIMOS and other optical spectrographs have difficulty in obtaining secure redshifts. The redshift and magnitude distibutions of the uncalibrated cells are summarized in Figure~\ref{figure:uncalibrated}. To fully complete the coverage would require deep spectroscopy on low sky density sources. Upcoming wide-area multiobject fiber spectrographs (e.g., the Prime Focus Spectrograph on Subaru; \citealp{Tamura16}) may be well-suited for this.

\begin{figure*}[htb]
\centering
      \includegraphics[width=.98\textwidth]{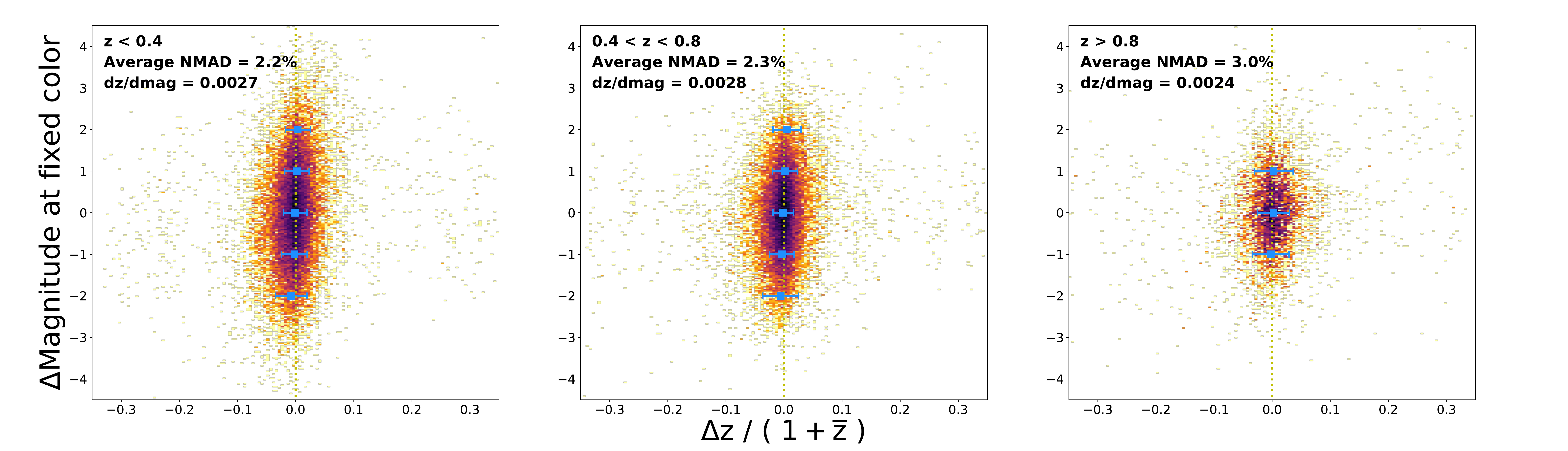} 
  \caption{The same plot as Figure~\ref{figure:mag_prior}, but this time splitting by redshift. The scatter and slope of the relation are virtually unchanged in different redshift bins. The weak but significant variation of redshift with magnitude at fixed color appears to be intrinsic, likely explained by subtle variations in color with mass/metallicity counterbalanced by slightly different redshift.}
      \label{figure:mag_prior_zbins}
\end{figure*}

\subsection{Magnitude dependence}

Important for \phz\ calibration is the extent to which the magnitude of a galaxy carries additional information about its redshift, relative to its observed colors. This is particularly relevant if it is found that the galaxies successfully targeted with spectroscopy are systematically brighter than average at a fixed point in the survey color space.  

We performed a test based on our full spectroscopic calibration sample to address this issue, comprising spectra from surveys such as zCOSMOS \citep{Lilly07}, DEEP2 \citep{Newman13}, and VVDS \citep{LeFevre15}, among others, in addition to the C3R2 spectra. There are many color cells in the SOM that contain multiple galaxies with high-confidence spectroscopic redshifts when considering all of these surveys. Therefore we take every \emph{unique pair} of spectroscopic galaxies at fixed color in our sample, and examine the correlation between their difference in magnitude and difference in redshift. The result is shown in Figure~\ref{figure:mag_prior}.

The scatter in $\Delta z /(1+\bar{z})$ in this figure tell us the uncertainty in the redshift at fixed color, both due to inherent uncertainty in the \pz\ relationship itself and photometric uncertainty. This value is $\sim$2.3\%, in agreement with the scatter we infer for the SOM-based photometric redshifts. More importantly, we find a very weak (but non-zero) correlation between $\Delta$mag and $\Delta z / (1+\bar{z})$ at fixed \e/\w\ color. The relationship has a slope of 0.003~mag$^{-1}$. While small, this relation is at a level that may be of concern for cosmology. Further detailed analysis would be required to understand how best to calibrate for the effect. However, the relation appears stable across redshift, as illustrated in Figure~\ref{figure:mag_prior_zbins}. Moreover, we tested whether it can be ascribed to small color variations of galaxies within a given SOM cell. This does not seem to be the explanation, as we find that the colors do not seem to systematically vary with galaxy brightness at a fixed SOM cell. We are investigating this issue using simulations and it will be explored more in a subsequent paper.

\subsection{Cosmic variance}

Because the calibration fields used by C3R2 are relatively small ($\sim$1-2~deg$^2$), cosmic variance will imprint a different \nz\ in each field than would be measured in wide-area surveys. This effect may impact the redshift calibration, and certainly would bias any \nz\ estimate drawn only from these fields, in comparison with the true \nz\ for a wide-area survey. 

Two critical questions are: (1) is the \pz\ relation itself unchanged across fields, regardless of variations in the relative numbers of galaxies at fixed color due to cosmic variance?, and (2) is the SOM generated on a handful of deep fields sufficiently spanning the color space of galaxies? 

\begin{figure*}[htb]
\centering
  \begin{tabular}{@{}cc@{}}
      \includegraphics[width=.4\textwidth]{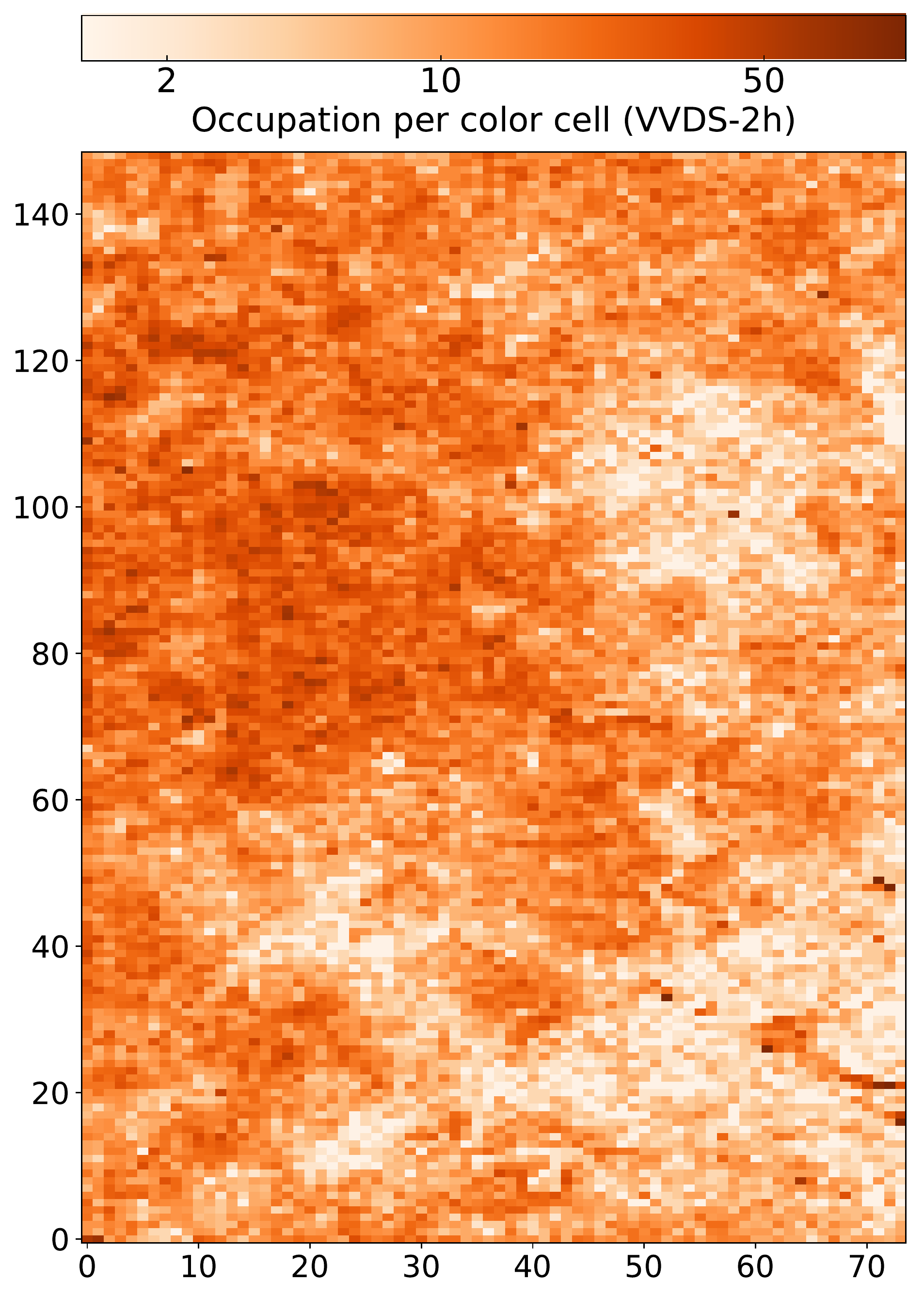} &
    \includegraphics[width=.4\textwidth]{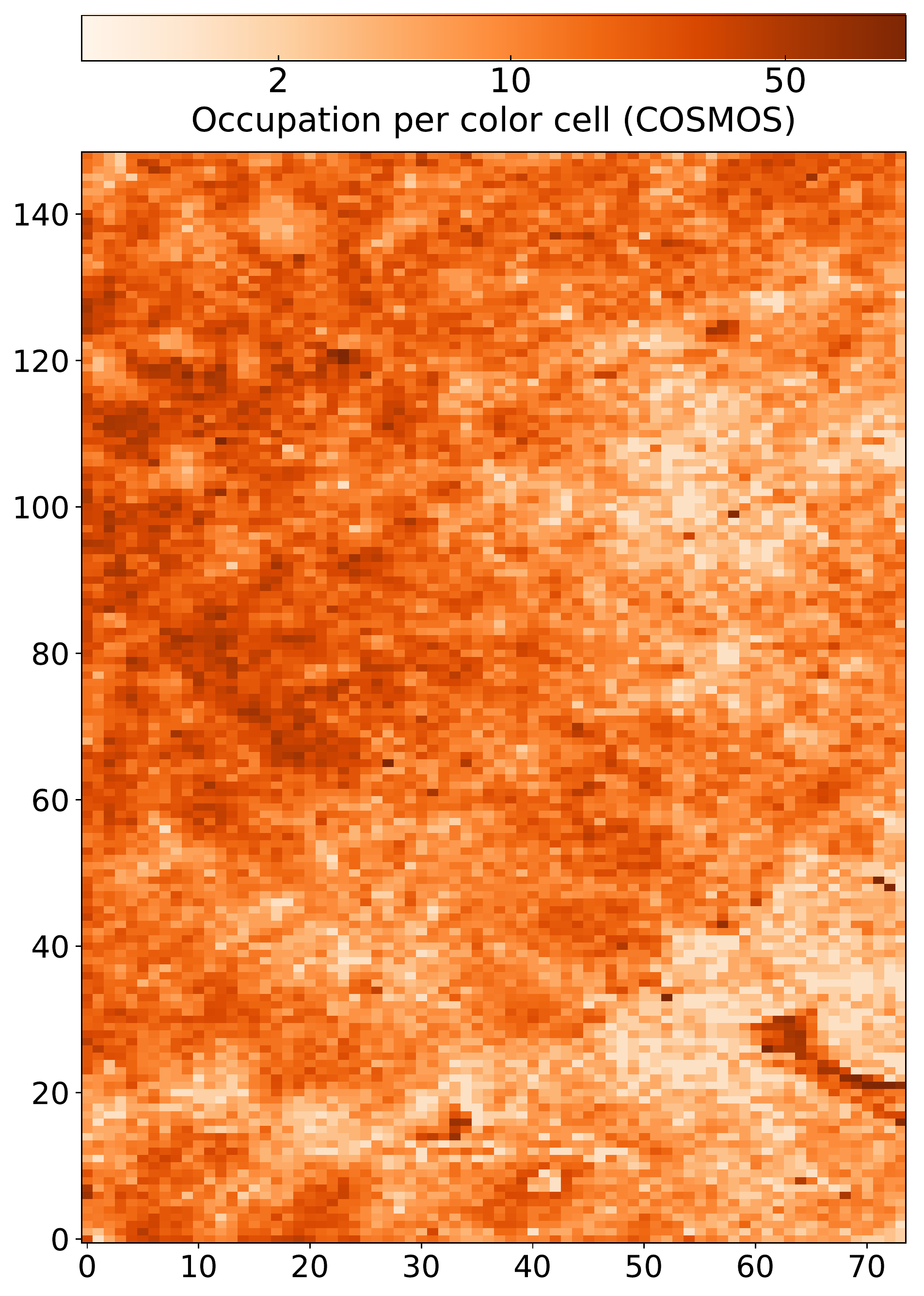} 
    \\    
  \end{tabular}
  \caption{Comparison of relative occupation density of galaxies in color space in COSMOS versus VVDS-2h. Variation of the fields is evident, and the median relative difference in occupation for a given cell of the color space is $\sim$50\%. If, however, we rebin the map in 5$\times$5 cells, the variance is much lower; at a median value of $\sim$25\% per color cell. Overall this illustrates the general similarity in frequency of colors in different fields, while also highlighting the impact of cosmic variance.}
      \label{figure:variance}
\end{figure*}

We can get a sense of the cosmic variance in the fields by examining the density field in color space, \rhoc, in VVDS-2h and COSMOS separately. As Figure~\ref{figure:variance} shows, roughly the same global density is found in each field, but with variations indicative of the different clustering in the fields. A simple assessment of whether the \pz\ relation is affected by this variation can be made by calibrating the relation with COSMOS only, then applying that solution to VVDS-2h. We performed this analysis and found a largely unchanged redshift results for the VVDS-2h sources; in other words, regardless of clustering, at fixed color the relation appears to be stable across fields. A more detailed analysis of the effect of cosmic variance on the calibration of \pz\ will be the subject of a future paper.

\section{Summary}
We have presented 3171 new redshifts obtained by the C3R2 survey. We have shown that the redshift calibration strategy for \e\ undertaken by C3R2 is making excellent progress, although it hinges strongly on obtaining highly homogeneous imaging across large deep fields. More extensive tests are required to ensure that we will meet the cosmology requirements. A key question is the extent to which cosmic variance will limit the utility of the redshift results in the deep fields when applying them to the future wide surveys. We have also shown an empirical relation between magnitude difference and redshift at fixed \e/\w\ color. This relation may prove important for extending the spectroscopic calibration of the \pz\ relation to deeper \w\ and LSST sources for which obtaining spectroscopy will remain challenging into the 2020s.

Our results illustrate that there is a strong, mostly non-degenerate relationship between observed galaxy optical-through-near-IR colors and redshift, and therefore uncovering this relation directly via targeted spectroscopy is a reliable method to calibrate redshifts for the dark energy experiments to be conducted over the next decade. Moreover, the quality of the redshift solution derived using the median multi-band photometric redshifts at each point in the \e\/\w\ color space suggests that deep surveys overlapping wide area surveys can be invaluable, particularly if they include more photometric bands than the wide survey. We emphasize that C3R2 is filling in regions of galaxy parameter space that are underrepresented in existing spectroscopic surveys. However, those surveys collectively have contributed hugely to mapping the galaxy color-redshift relation. 

A key goal for the community interested in calibrating photometric redshifts for weak lensing cosmology should be to develop a homogenized database of \emph{all} 1D/2D spectra taken to date from which all surveys can benefit. As we have shown, care is needed to avoid calibrating the \pz\ relation with ambiguous sources, nearby galaxies in projection, etc. These can be a sizable fraction particularly at the depths of the upcoming surveys, and thus a carefully curated database of reliable and well-vetted spectroscopy is needed. 

\acknowledgments
We thank the anonymous referee for a thorough and constructive report that improved this manuscript. The research was carried out at the Jet Propulsion Laboratory, California Institute of Technology, under a contract with the National Aeronautics and Space Administration. D.M., D.S., P.C., and J.R. acknowledge support by NASA ROSES grant 12-EUCLID12-0004. D.M. acknowledges support for this work from a NASA Postdoctoral Program Fellowship. This work was enabled by a NASA Keck grant. 







\smallskip

\smallskip
\copyright 2018.  All rights reserved.

\clearpage

\bibliographystyle{apj.bst}
\bibliography{new.ms.bib}

\clearpage

\appendix

\scriptsize
\begin{deluxetable*}{lccccc}
\tablecaption{List of observed slitmasks.}
\tabletypesize{\footnotesize}
\tablehead{
\colhead{} &
\colhead{} &
\colhead{R.A.} &
\colhead{Dec.} &
\colhead{PA} &
\colhead{Exposure} \\
\colhead{Mask ID / Name} &
\colhead{Night} &
\colhead{(J2000)} &
\colhead{(J2000)} &
\colhead{(\deg)} &
\colhead{(s)} }
\startdata
16B-L027 / VVDS-m01     & N06-L &  2:24:54.5 &  $-$4:37:15 & 8.8 & 6$\times$1200  \\ 
16B-L028 / VVDS-m02    & N06-L &  2:24:52.8 &  $-$4:04:02 &  $-$20.1 & 6$\times$1200  \\ 
16B-L029 / VVDS-m03    & N06-L &  2:25:03.8 &  $-$4:07:43 & 158.5 & 3$\times$1200  \\ 
16B-D030 / VVDS-m03     & N07-D &  2:27:17.8 &  $-$4:40:03 & 90.0 & 6$\times$1200*  \\ 
16B-D031 / VVDS-m01   & N08-D &  2:27:18.0 &  $-$4:53:40 & 90.0 & 4$\times$1200*\footnote{3-hour mask abandoned during fifth exposure due to heavy cloud cover.  Re-observed on N11-D as mask 16B-D043.}  \\ 
16B-D032 / VVDS-m05    & N08-D &  2:27:22.4 &  $-$4:22:12 & 90.0 & 3$\times$1200*  \\ 
16B-D033 / VVDS-m04   & N08-D &  2:27:24.7 &  $-$4:28:54 & 90.0 & 6$\times$1200*  \\ 
16B-L034 / VVDS-m04     & N09-L &  2:27:11.0 &  $-$4:04:10 & 0.0 & 9$\times$1200  \\
16B-D035 / VVDS-m06    & N10-D &  2:27:16.2 &  $-$4:12:12 & 90.0 & 4$\times$1200  \\ 
16B-D036 / VVDS-m21    & N10-D &  2:24:49.8 &  $-$4:44:00 & 90.0 & 6$\times$1200  \\ 
16B-D037 / VVDS-m22    & N10-D &  2:24:50.5 &  $-$4:39:53 & 90.0 & 5$\times$1200  \\ 
16B-D038 / VVDS-m07    & N10-D+N12-D &  2:26:22.7 &  $-$4:04:00 & 90.0 & 2$\times$900 + 2$\times$1200  \\ 
16B-D039 / COSMOS-m11  & N10-D &  9:58:42.4 &  $+$2:37:37 & 90.0 & 4$\times$1200  \\ 
16B-D040 / COSMOS-m12  & N10-D &  9:58:42.8 &  $+$2:42:34 & 90.0 & 3$\times$1200  \\ 
16B-D041 / COSMOS-m13   & N10-D &  9:58:35.7 &  $+$2:45:47 & 90.0 & 2$\times$1200 + 900  \\ 
16B-D042 / VVDS-m08    & N11-D &  2:26:00.0 &  $-$4:11:55 & 90.0 & 3$\times$1200  \\ 
16B-D043 / VVDS-m01    & N11-D &  2:27:18.0 &  $-$4:53:40 & 90.0 & 9$\times$1200  \\ 
16B-D044 / VVDS-m23     & N11-D &  2:24:50.2 &  $-$4:36:00 & 90.0 & 5$\times$1200  \\ 
16B-D045 / COSMOS-m14  & N11-D &  9:59:55.3 &  $+$1:41:54 & 90.0 & 4$\times$1200  \\ 
16B-D046 / COSMOS-m21  & N11-D &  9:59:53.3 &  $+$2:10:00 & 90.0 & 7$\times$1200  \\ 
16B-D047 / VVDS-m09    & N12-D &  2:25:57.2 &  $-$4:20:18 & 90.0 & 2$\times$900  \\ 
16B-D048 / COSMOS-m16  & N12-D &  9:59:48.2 &  $+$1:49:05 & 90.0 & 3$\times$1200 \\ 
16B-D049 / VVDS-m10    & N15-D &  2:25:53.1 &  $-$4:28:36 & 90.0 & 2$\times$1200 + 1058  \\ 
16B-D050 / VVDS-m02    & N17-D &  2:27:20.3 &  $-$4:48:03 & 90.0 & 9$\times$1200  \\ 
16B-D051 / VVDS-m11    & N17-D &  2:26:01.1 &  $-$4:32:18 & 90.0 & 2$\times$1200 + 900  \\ 
16B-D052 / COSMOS-m15  & N17-D &  9:59:50.4 &  $+$1:46:00 & 90.0 & 4$\times$1200  \\ 
16B-D053 / COSMOS-m24  & N17-D &  9:59:52.4 &  $+$2:22:00 & 90.0 & 9$\times$1200  \\ 
16B-D054 / COSMOS-m17  & N17-D &  9:59:51.6 &  $+$1:54:32 & 90.0 & 4$\times$1200  \\ 
16B-D055 / VVDS-m12    & N18-D &  2:26:02.8 &  $-$4:36:12 & 90.0 & 3$\times$1200  \\ 
16B-D056 / VVDS-m24    & N18-D &  2:24:50.1 &  $-$4:32:24 & 90.0 & 5$\times$1200  \\ 
16B-D057 / COSMOS-m18  & N18-D &  9:59:56.5 &  $+$1:54:59 & 90.0 & 4$\times$1200  \\ 
16B-D058 / COSMOS-m22  & N18-D &  9:59:53.0 &  $+$2:14:00 & 90.0 & 6$\times$1200  \\ 
16B-D059 / COSMOS-m23  & N18-D &  9:59:52.4 &  $+$2:18:00 & 90.0 & 7$\times$1200  \\ 
16B-D060 / COSMOS-m27  & N19-D &  9:59:55.0 &  $+$2:35:00 & 90.0 & 5$\times$1200  \\ 
16B-D061 / COSMOS-m25   & N19-D & 10:00:17.3 &  $+$2:25:56 & 90.0 & 9$\times$1200  \\ 
16B-D062 / COSMOS-m28  & N19-D &  9:59:57.3 &  $+$2:37:54 & 90.0 & 3$\times$1200  \\ 
16B-D063 / VVDS-m13    & N20-D &  2:25:59.6 &  $-$4:40:09 & 90.0 & 3$\times$1200 \\ 
16B-D064 / VVDS-m14    & N20-D &  2:25:59.2 &  $-$4:43:50 & 90.0 & 2$\times$1150 + 1250 \\ 
16B-D065 / COSMOS-m19  & N20-D &  9:59:53.0 &  $+$2:01:56 & 90.0 & 3$\times$1200  \\ 
16B-D066 / COSMOS-6h   & N20-D+N22-D & 10:02:34.6 &  $+$2:48:01 & 90.0 & 18$\times$1200 + 1600  \\ 
16B-D067 / COSMOS-m32  & N20-D &  9:59:56.8 &  $+$2:45:12 & 90.0 & 3$\times$1200  \\ 
16B-L068 / VVDS-m7     & N21-L &  2:25:55.2 &  $-$4:13:42 & 148.4 & 6$\times$1200  \\ 
16B-L069 / VVDS-m5      & N21-L &  2:26:26.7 &  $-$4:42:45 & 1.7 & 3$\times$1200  \\ 
16B-L070 / COSMOS-m10  & N21-L & 10:01:01.7 &  $+$2:41:05 & 101.0 & 4$\times$1200 + 900  \\ 
16B-L071 / COSMOS-m13  & N21-L &  9:59:05.8 &  $+$1:45:13 & 178.4 & 9$\times$1200   \\ 
16B-L072 / COSMOS-m11  & N21-L & 10:00:00.2 &  $+$2:04:25 & 31.0 & 5$\times$1200  \\ 
16B-L073 / VVDS-m25    & N22-D &  2:24:49.8 &  $-$4:28:04 & 90.0 & 5$\times$1200 + 1150  \\ 
16B-D074 / COSMOS-m33  & N22-D & 10:01:07.8 &  $+$2:45:12 & 90.0 & 3$\times$1200  \\ 
16B-D075 / COSMOS-m34  & N22-D & 10:01:03.0 &  $+$2:41:00 & 90.0 & 3$\times$1200 + 1150  \\ 
16B-D076 / COSMOS-m39  & N22-D & 10:01:13.8 &  $+$2:21:02 & 90.0 & 7$\times$1200  \\ 
17A-D077 / COSMOS-5h1  & N23-D+N24-D & 10:01:41.9 & $+$2:04:24  & 90.0 & 15$\times$1200  \\ 
17A-D078 / EGS-4h1     & N23-D+N24-D & 14:18:00.1 & $+$52:46:00  & 90.0 & 13$\times$1200  \\ 
17A-D079 / COSMOS-1h1  & N24-D & 10:00:19.2 & $+$2:15:21  & 90.0 & 4$\times$1200  \\ 
17A-D080 / COSMOS-1h2  & N24-D & 9:59:59.4 & $+$2:02:44   & 90.0 & 3$\times$1200  \\ 
17A-D081 / EGS-4h2   & N24-D+N26-D & 14:17:59.8 & $+$52:51:00  & 90.0 & 15800\footnote{4$\times$1200 + 5$\times$1800 + 2000}  \\ 
17A-L082 / COS-lriD/COS-5hr1 & N25-L & 10:01:43.8 & $+$2:23:31  & 129.9 & 15$\times$1200  \\ 
17A-L083 / EGS-lriE/EGS-4hr1 & N25-L & 14:19:35.9 & $+$53:05:23  & $-$131.6 & 11$\times$1200  \\ 
17A-D084 / COSMOS-5h3  & N26-D & 9:59:46.8 & $+$2:19:00  & 90.0 & 13$\times$1200  \\ 
17A-D084 / EGS-1h5   & N26-D & 14:18:00.3 & $+$52:36:00  & $-$147.1 & 1$\times$1800  \\ 
17A-L085 / COS-lriE/COS-5hr2 & N27-L & 10:01:29.1 & $+$2:21:34  & 90.0 & 13$\times$1200  \\ 
17A-L086 / EGS-lriF/EGS-4hr2 & N27-L & 14:18:07.4 & $+$53:02:06  & $-$171.5 & 14$\times$1200  \\ 
17A-M087 / COSMOS-H-2hr3 & N28-M & 10:01:49.8 & $+$2:36:02 & $-$12.0 & 60$\times$120  \\ 
17A-M088 / EGS-H-2hr3   & N28-M & 14:17:30.9 & $+$52:49:33 & 35.0 & 60$\times$120  \\ 
17A-M089 / EGS-H-2hr4   & N28-M & 14:19:00.4 & $+$53:05:14 & 5.0 & 60$\times$120  \\ 
17A-M090 / COSMOS-H-2hr1 & N29-M &  9:58:33.7 & $+$1:50:59 & 20.0 & 60$\times$120  \\ 
17A-M091 / COSMOS-H-1hr6 & N29-M & 10:01:29.9 & $+$1:52:23 & 30.0 & 24$\times$120  \\ 
17A-M092 / EGS-H-2hr5   & N29-M & 14:17:01.8 & $+$52:56:32 & 35.0 & 60$\times$120  \\ 
17A-M093 / EGS-H-2hr6    & N29-M & 14:20:11.1 & $+$53:00:10 & 30.0 & 60$\times$120  
\enddata
\label{table:slitmasks}
\tablecomments{`Night' column refers to observing code in second
column of Table~1: night number, followed by letter indicating
instrument used (D -- DEIMOS, L -- LRIS, M -- MOSFIRE).  R.A. and
Dec.  refer to the mask center.  Final column gives total number
of slitlets in mask, total number of high-quality (Q=4) redshifts
measured, and the number of serendipitous sources with high-quality
redshifts (Q=4). * - Taken under poor conditions. See notes.}
\end{deluxetable*}
\normalsize


\begin{deluxetable*}{cccccccccc}
\tablecaption{C3R2 spectroscopic redshift results \label{table:catalog}}
\tablehead{
\colhead{ID} & 
\colhead{R.A.} & 
\colhead{Dec.} & 
\colhead{Mask} & 
\colhead{Slit} & 
\colhead{$i$~mag} & 
\colhead{$z$} & 
\colhead{Qual.} & 
\colhead{Instr.} & 
\colhead{File} \\
\colhead{} & 
\colhead{(J2000)} & 
\colhead{(J2000)} & 
\colhead{} & 
\colhead{} & 
\colhead{} & 
\colhead{} & 
\colhead{} & 
\colhead{} & 
\colhead{} 
}
\startdata
UDS-3583 & 02:17:30.65 & -05:15:24.4 & UDS-m1n1 & 1 & 23.4 & 0.7877 & 4.0 & DEIMOS & spec1d.u-m1n1.001.UDS-3583.fits \\
UDS-10246 & 02:17:17.55 & -05:13:06.9 & UDS-m1n1 & 2 & 23.9 & 0.8028 & 4.0 & DEIMOS & spec1d.u-m1n1.002.UDS-10246.fits \\
UDS-767 & 02:17:59.05 & -05:16:21.2 & UDS-m1n1 & 3 & 25.0 & 0.5558 & 4.0 & DEIMOS & spec1d.u-m1n1.003.UDS-767.fits \\
UDS-7109 & 02:17:00.35 & -05:14:15.4 & UDS-m1n1 & 4 & 23.6 & 1.0314 & 3.0 & DEIMOS & spec1d.u-m1n1.004.UDS-7109.fits \\
UDS-2276 & 02:17:52.83 & -05:15:55.2 & UDS-m1n1 & 5 & 22.9 & 0.9388 & 4.0 & DEIMOS & spec1d.u-m1n1.005.UDS-2276.fits \\
UDS-8536 & 02:17:53.37 & -05:13:40.3 & UDS-m1n1 & 6 & 24.7 & 0.8619 & 4.0 & DEIMOS & spec1d.u-m1n1.006.UDS-8536.fits \\
UDS-9784 & 02:17:56.80 & -05:13:15.9 & UDS-m1n1 & 9 & 23.6 & 0.8533 & 4.0 & DEIMOS & spec1d.u-m1n1.009.UDS-9784.fits \\
UDS-10739 & 02:17:44.88 & -05:12:58.7 & UDS-m1n1 & 10 & 23.2 & 1.0594 & 4.0 & DEIMOS & spec1d.u-m1n1.010.UDS-10739.fits \\
UDS-9730 & 02:17:32.64 & -05:13:17.4 & UDS-m1n1 & 12 & 23.8 & 1.0949 & 4.0 & DEIMOS & spec1d.u-m1n1.012.UDS-9730.fits \\
\ldots 
\enddata
\tablecomments{Table \ref{table:catalog} is published in its entirety in the machine-readable format. A portion is shown here for guidance regarding its form and content. More detailed information is available in the header of the machine-readable table.}
\end{deluxetable*}

\end{document}